\documentclass[onecolumn,12pt]{IEEEtran} 

\usepackage{textcomp}
\usepackage[latin9]{inputenc}
\usepackage{float}
\usepackage{tikz}
\usepackage{amsmath}
\usepackage{amssymb}

\usepackage{diagbox}
\usepackage{cite}

\usepackage{amsthm}
\usepackage{graphicx}
\usepackage[acronym]{glossaries}
\usepackage{algorithm2e}
\usepackage{url}

\makeatletter

\floatstyle{ruled}
\newfloat{lyxalgorithm}{tbp}{loa}
\providecommand{\algorithmname}{Algorithm}
\floatname{lyxalgorithm}{\protect\algorithmname}

\SetAlgoLined
\LinesNumbered
\RestyleAlgo{ruled} 

\theoremstyle{plain}
\newtheorem{thm}{\protect\theoremname}
\theoremstyle{plain}
\newtheorem{prop}[thm]{\protect\propositionname}
\ifx\proof\undefined
\newenvironment{proof}[1][\protect\proofname]{\par
	\normalfont\topsep6\p@\@plus6\p@\relax
	\trivlist
	\itemindent\parindent
	\item[\hskip\labelsep\scshape #1]\ignorespaces
}{%
	\endtrivlist\@endpefalse
}
\providecommand{\proofname}{Proof}
\fi
\theoremstyle{plain}
\newtheorem{cor}[thm]{\protect\corollaryname}

\makeatother

\providecommand{\corollaryname}{Corollary}
\providecommand{\propositionname}{Proposition}
\providecommand{\theoremname}{Theorem}

\newacronym{ai}{AI}{Artificial Intelligence}
\newacronym{ann}{ANN}{Artificial Neural Networks}
\newacronym{arma}{ARMA}{Auto-Regressive Moving-Average}
\newacronym{ar}{AR}{Autoregressive}
\newacronym{arp}{ARP}{Autoregressive Process}
\newacronym{asbc}{ASBC}{Adaptive Subband Compression}
\newacronym{asdu}{ASDU}{Application Service Data Unit}
\newacronym{bf}{BFS}{Brute Force Search}
\newacronym{bpc}{BPC}{Bit-Plane Coding}
\newacronym{bpd}{BPD}{Bit-Plane Decoding}
\newacronym{bps}{bps}{Bit Per Sample}
\newacronym{bp}{BP}{Basis Pursuit}
\newacronym{cabac}{CABAC}{Context-Adaptive Binary Arithmetic Coding}
\newacronym{cdf}{CDF}{Cumulative Distribution Function}
\newacronym{chdm}{CHDM}{Cyclical High-Order Delta Modulation}
\newacronym{cr}{CR}{Compression Ratio}
\newacronym{cs}{CS}{Compressive Sensing}
\newacronym{cnes}{NSOC}{National System Operations Center}
\newacronym{cnn}{CNN}{Convolutional Neural Network}
\newacronym{dm}{DM}{Distortion Model}
\newacronym{dct}{DCT}{Discrete Cosine Transform}
\newacronym{dft}{DFT}{Discrete Fourier Transform}
\newacronym{dfrs}{DFRS}{Digital Fault Recorders}
\newacronym{dwt}{DWT}{Discrete Wavelet Transform}
\newacronym{ec}{EC}{Entropy Coding}
\newacronym{ed}{ED}{Entropy Decoding}
\newacronym{es}{ES}{Exhaustive Search}
\newacronym{ezw}{EZW}{Embedded Zerotree Wavelet}
\newacronym{fft}{FFT}{Fast Fourier Transform}
\newacronym{ga}{GA}{Genetic Algorithms}
\newacronym{gan}{GAN}{Generative Adversarial Network}
\newacronym{gdn}{GDN}{generalized Divisive Normalization}
\newacronym{gss}{GSS}{Golden Section Search}
\newacronym{gt}{GT}{Gabor Transform}
\newacronym{gwt}{GWT}{Gabor-Wigner Transform}
\newacronym{hht}{HHT}{Hilbert-Huang Transform}
\newacronym{ht}{HT}{Hilbert Transform}
\newacronym{ica}{ICA}{Independent Component Analysis}
\newacronym{iec}{IEC}{International Electrotechnical Commission}
\newacronym{ieee}{IEEE}{Institute of Electrical and Electronics Engineers}
\newacronym{jpeg}{JPEG}{Joint Photographic Experts Group}
\newacronym{kl}{KL}{Kullback-Leibler}
\newacronym{lzma}{LZMA}{Lempel-Ziv-Markov chain Algorithm}
\newacronym{lzw}{LZW}{Lempel Ziv Welch}
\newacronym{mdl}{MDL}{Minimum Description Length}
\newacronym{mmc}{MMC}{Multiple-Model Coding}
\newacronym{mp}{MP}{Matching Pursuit}
\newacronym{mrd}{MRD}{Model Rate Distortion}
\newacronym{mse}{MSE}{Mean Square Error}
\newacronym{nse}{NSE}{Neural Shape Estimator}
\newacronym{ohd}{OHD}{Overcomplete Hybrid Dictionaries}
\newacronym{omp}{OMP}{Orthogonal Matching Pursuit}
\newacronym{pacf}{PACF}{Partial Autocorrelation Function}
\newacronym{pca}{PCA}{Principal Component Analysis}
\newacronym{pc}{PC}{Principal Component}
\newacronym{pdc}{PDC}{Phasor Data Concentrator}
\newacronym{pll}{PLL}{Phased-Locked Loop}
\newacronym{pmu}{PMU}{Phasor Measurement Unit}
\newacronym{pow}{POW}{Point-Of-Wave}
\newacronym{q}{Q}{Quantization}
\newacronym{rd}{RD}{Rate-Distortion} 
\newacronym{re}{RE}{Renewable Energy} 
\newacronym{rte}{RTE}{Réseau de Transport d'Électricité}
\newacronym{scada}{SCADA}{Supervisory Control And Data Acquisition}
\newacronym{snr}{SNR}{Signal to Noise Ratio}
\newacronym{st}{ST}{S-Transform}
\newacronym{stft}{STFT}{Short-Time Fourier Transform}
\newacronym{sttp}{STTP}{Streaming Telemetry Transport Protocol}
\newacronym{sv}{SV}{Sample Value}
\newacronym{qp}{QP}{Quantization Parameter}
\newacronym{tcp}{TCP}{Transmission Control Protocol}
\newacronym{udp}{UDP}{User Datagram Protocol}
\newacronym{uhd}{UHD}{Ultra High-Density}
\newacronym{vae}{VAE}{Variational Auto Encoder}
\newacronym{vaes}{VAEs}{Variational Autoencoders}
\newacronym{vcs}{VCS}{Voltage and Current Signal}
\newacronym{wan}{WAN}{Wide Area Network}
\newacronym{wdf}{WDF}{Wigner Distribution Function}
\newacronym{wsqm}{WSQM}{Wavelet Spectral Quantization Models}
\newacronym{wt}{WT}{Wavelet Transform}

\begin{document}
	
	\title{Reduced-Complexity \\Model Selection and Rate Allocation\\
		for Multiple-Model Electrical Signal Compression}
	
	\author{Corentin Presv\^ots, Michel Kieffer,~\IEEEmembership{Senior Member,~IEEE,} Thibault Prevost
		}
	
	
	
	\maketitle
	
	\begin{abstract}
		This paper adapts a \gls{mmc} approach for sampled electrical signal waveforms to satisfy reconstructed signal quality constraints. The baseline \gls{mmc} approach consists of two stages processing vectors of \gls{vcs} of constant size and producing bitstreams of constant rate but varying quality. In the proposed approach, the parametric model and the rate allocated to the first stage, as well as the residual compression method of the second stage and its associated rate, are jointly optimized to achieve a target distortion of the reconstructed signal. Three approaches are proposed. An exhaustive search serves as a baseline for comparison. Then, an approach involving a Golden Section search is exploited to determine the rate of the first stage with reduced complexity. Finally, rate-distortion models of the compression efficiency for each model in the first stage are employed to obtain a subset of promising models in the first stage and reduced-size search intervals for the rate selection in both stages. Simulation results demonstrate that the proposed reduced-complexity \gls{mmc} approach reduces the rate for a given distortion constraint compared to state-of-the-art solutions for \gls{vcs} with equivalent complexity.
	\end{abstract}
	
	\begin{IEEEkeywords}
		Compression, low latency, model-based coding, rate-distortion optimization, electrical waveform compression
	\end{IEEEkeywords}

\section{Introduction}

\label{sec:Introduction}

Integrating renewable energy sources within the electrical grid introduces complex grid dynamics, as highlighted by \cite{Gu2023}. A more efficient, reduced-latency control of the grid may be necessary. Such control requires transmitting large volumes of data from substations to higher-level control centers. To meet the rate constraints of communication networks, it is essential to develop efficient compression techniques tailored to the specific characteristics of electrical signals. Currently,
\glspl{pmu} are the most commonly deployed devices for acquiring and compressing electrical signals. \glspl{pmu}, however, are unable to represent rapid transients accurately. \gls{vcs} are better suited to detect and localize faults in the network or reduced-latency control tasks. Several techniques have been developed to compress \gls{vcs}, as discussed in Section~\ref{sec:Related work}.

This work extends the approach in \cite{MMC_DCC}, which introduced
a two-stage \gls{mmc} method for \gls{vcs}. In the first stage, several parametric models are compared to obtain a coarse, low-bit-rate representation of the samples. In the second stage, various residual
compression techniques are evaluated to minimize distortion. The selection of the optimal model in the first stage and the residual compression technique in the second stage, as well as the bit allocation between the two stages, are optimized considering a total bit budget constraint. This optimization may be complex and incompatible with real-time constraints. Moreover, imposing a bit budget constraint leads to variations in the reconstruction quality of the \gls{vcs}.

This paper aims at determining the minimum total bit budget required to satisfy some maximum distortion constraint when \gls{vcs} are compressed with the \gls{mmc} approach in \cite{MMC_DCC}. This approach ensures a target reconstruction quality, but requires a proper adaptation of the bit budget for each encoded group of the \gls{vcs}, which increases the complexity of the \gls{mmc} approach compared to the fixed-rate variant. Three methods are proposed. The first method involves an \gls{es} within some progressively refined interval to determine the bit allocation for the first stage. The second method involves the \gls{gss} within the search interval. Finally, the third method leverages \gls{rd} models to get an initial estimate of the bit allocation between the two stages for each considered model and to select a subset of promising candidate models and reduced-size search intervals for the rate. The optimal allocation between stages is then determined by applying one of the first two methods.

The remainder of this paper is organized as follows. Section~\ref{sec:Related work} reviews some related work. Section~\ref{sec:Overview MMC} briefly summarizes the \gls{mmc} approach of \cite{MMC_DCC} and introduces the problem of optimizing the total bit budget, the model in the first stage, the compression method in the second stage, and the bit allocation between the two stages to meet some maximum distortion constraint. The methods for solving this problem are discussed in Section~\ref{sec:Proposed solutions}. Section~\ref{sec:Distortion models for both stages} discusses a solution using \gls{rd} models. Simulation results are presented in Section~\ref{sec:Simulations examples}. Section~\ref{sec:Conclusion}
concludes this paper.

\section{Related work\label{sec:Related work}}

In lossy data compression, a compromise has to be found between the
rate at which data are encoded, \emph{i.e.}, the number of bits used
to represent the compressed data, and the distortion of the reconstructed
version of the data. Section~\ref{subsec:En-compression-d'images} briefly
recalls the methods used in image and video compression to reach a
target rate-distortion compromise. Then, Section~\ref{subsec:Simple-Approach-to}
recalls techniques for compressing electrical signals, some focusing
on a target rate, others on a target distortion. Finally, Section~\ref{subsec:More-Complex-Approach}
presents various two-stage compression methods, where an initial coarse
compression is refined in a second stage to improve the rate-distortion
balance. These methods involve higher computational costs, particularly
to adjust the bit allocation between the two stages while satisfying
a target rate or distortion constraint. This section also explores
strategies proposed in the literature to balance computational complexity
with rate or distortion objectives.

\subsection{Satisfying a rate or a distortion constraint in image or video coding\label{subsec:En-compression-d'images}}

In real-time streaming, particularly over wireless networks, the available
bandwidth is often constrained and time varying. Encoding Rate Control
(RC) methods are then employed to adjust the compression rate to minimize
distortion and avoid receiver buffer depletion. This is typically
done by tuning the value of the Quantization Parameters (QPs) employed
after transform coding of the prediction residuals. Instead of testing
all combinations of QPs, rate and distortion models are exploited
to evaluate the trade-off between rate and distortion as a function
of characteristics of the encoded stream or of the coding parameters,
such as the QPs.

Rate and distortion models fall into two broad categories: those assuming
that data blocks serving as the basis for the compression process
such as frames or Coding Units (CU) are independent, and those accounting
for inter-dependencies.

For the first category, R-QP models predict the rate as a function
of the quantization parameter QP by estimating the distortion between
a coding unit and its predicted version, which allows the model to
adapt the predicted rate according to the signal content. Commonly-used
distortion metrics include the Mean Absolute Difference (MAD) \cite{ding1996rate,choi2013pixel},
the Sum of Absolute Difference (SAD) \cite{ma2005rate,li2014inter},
and the \gls{mse} \cite{aklouf2021interframe}.. With R-$\rho$ models
\cite{liu2010low}, the rate is predicted based on the percentage
of zero-valued transformed coefficients. Finally, R-$\lambda$ models
use a Lagrangian multiplier $\lambda$ to control the rate-distortion
trade-off \cite{tang2019generalized}. A parametric R-D model is fitted
from videos encoded at different QPs, and $\lambda$ is defined as
the negative slope of the R-D curve. The corresponding QP for a given
$\lambda$ approximately minimizes $D+\lambda R$, achieving the desired
balance between rate and distortion. Experiments in \cite{li2014lambda,li2016lambda}
show that R-$\lambda$-based rate control algorithms outperform those
based on R-QP or R-$\rho$ models. 

In the second category, methods like those in \cite{tang2019generalized,yang2006rate}
predict the rate using the standard deviation of motion-compensated
residuals. Unlike R-$\rho$ models, where MAD, SAD, or \gls{mse}
are computed solely from the local characteristics of a CU, these
models account for the propagation between blocks of the prediction
errors. Other approaches use cubic models \cite{lin1998bit,zhang2003constant}
to predict the rate based on the values of QP of both the current
and reference frames. In \cite{aklouf2021interframe}, an R-D-QP model
is considered where the rate is predicted as a function of the distortion
of the reference frame and the average QP of the current frame. Deep
learning-based methods are also used to estimate the reconstruction
quality at a given rate using previous images and the current image,
allowing the frame reconstruction quality to be adjusted via QP \cite{huang2018qarc,zhou2020rate}.

Scalable coders generate a single bitstream from which reconstructions
with different quality levels are obtained, depending on the proportion
of bits taken from the bitstream \cite{schwarz2007overview,boyce2015overview}.
Coarse-Grain Scalability (CGS) \cite{zhao2015ssim,gupta2012h} refines
the prediction residuals in discrete steps or layers, each providing
an improvement in quality. Fine Grain Scalability (FGS) \cite{skodras2001jpeg}
progressively refines the prediction residuals with finer increments,
allowing for a more flexible adjustment of quality based on the available
rate. Historical examples of scalable coding methods dedicated to
image coding are \gls{ezw} \cite{shapiro_embedded_1993}, Layered
Zero Coding (LZC) \cite{taubman1994multirate}, Set Partitioning in
Hierarchical Trees (SPIHT) \cite{said1996new}, or Embedded Block
Coding with Optimized Truncation of the embedded bit-streams (EBCOT)
\cite{taubman2000high}. Recent scalable extensions of video coders
are Scalable High Efficiency Video Coding (SHEVC) \cite{nguyen2020scalable},
Scalable Video Technology (SVT) for AOMedia Video 1 (AV1) \cite{kossentini2020svt},
or the Low Complexity Enhancement Video Coding (LCEVC) \cite{battista2022overview}.
Nevertheless, generating a scalable bitstream usually results in some
loss in compression performance compared to a non-scalable approach
tuned using an R-D model \cite{schwarz2007overview}.

In recently-developed neural compression approaches \cite{balle_end--end_2017,balle_variational_2018,lee_context-adaptive_2019,kong2023mixture,aggarwal2023neural},
a trainable autoencoder architecture is optimized by taking into account
both quantization and entropy coding. The loss function is the rate-distortion
trade-off $R+\lambda D$ for a given regularization parameter $\lambda$,
but there is no strict constraint on the rate or distortion and several
pre-trained models for different values of $\lambda$ are required
to handle different target rates. To address this issue, coarse-grain
scalable learning-based coding schemes have been proposed \cite{mei2021learning,zhang2023exploring,fu2024learned}.
A base-resolution layer provides a coarse reconstruction of the data,
followed by several enhancement layers. To reduce the correlation
between layers, each enhancement layer predicts the residuals during
training.

\subsection{Baseline approaches for rate or distortion constraint satisfaction
in electrical signal compression\label{subsec:Simple-Approach-to}}

\subsubsection{Parametric coding}

Parametric coding approaches represent signals with one or few parametric
models including sum of sinusoids or damped sinusoids \cite{santoso1997power,chung2000electric,gu2003bridge,ribeiro2007classification,bollen2009bridging,khokhar2017new,yan2023review}
to exploit the physical properties of signals to compress. Phasor
Measurement Units (PMUs) use a sinusoidal model characterized by its
amplitude, frequency, and phase. In the standard \cite{IEEE_C37_118_2},
each of these three parameters is encoded using 12 bytes. Depending
on the acquisition frequency, the volume of data generated can be
large. For example, one \glspl{pmu} collecting 20 measurements at
30 Hz sampling rate generates over 500 MBytes of data per day \cite{klump_lossless_2010}.

The compression approaches for non-stationary signals in power systems
have recently shifted from traditional phasor-based approaches towards
methods that leverage more sophisticated parametric models. The classic
stationary phasor representation is often insufficient to capture
the broadband dynamics induced by modern grid conditions, such as
frequency ramps, amplitude modulations, and step changes in renewable-dominated
networks \cite{paolone_fundamentals_2020}.

The sinusoidal model used in the PMUs has been extended by decomposing
the signal into a sum of parametric damped sinusoids \cite{zygarlicki_data_2006,zygarlicki_reduced_2010}
each of which is represented by four parameters: amplitude, frequency,
phase, and damping factor. Similarly, \cite{Lovisolo_Efficient_coherent_adaptive_representations2005,lovisolo_modeling_2007}
decomposes the signal into a sum of piecewise damped sinusoids, requiring
two additional parameters per model to define the window size for
each damped sinusoid. In \cite{tcheou_optimum_2007,tcheou2012far},
the signal is also decomposed as a sum of damped sinusoids, gathered
in a parametric dictionary. The minimum number of dictionary elements
(atoms) is selected so as to satisfy a reconstruction quality constraint.
The atom parameters are quantized with several quantizers, each of
which defines a dictionary of quantized parameters. Therefore, testing
different quantizers is equivalent to testing different dictionaries.
For every candidate dictionary, the rate--distortion trade-off is
evaluated, and the one that minimizes bitrate while respecting the
distortion constraint is selected. 

Extending these ideas, \cite{karpilow_characterization_2021,karpilow2024functional}
introduce a dictionary-based approach that uses a set of predefined
parametric models such as amplitude modulations, frequency ramps,
phase modulations, amplitude steps, and phase steps, to model typical
grid disturbances. In \cite{yaghoobi2009parametric}, the dictionary
is iteratively constructed using equiangular tight frames with minimal
coherence to enhance signal representation and separation. Nevertheless,
no explicit quantization or entropy coding is applied. Compression
is achieved by retaining only a small subset of dictionary elements
to represent the original signal.

\subsubsection{Transform coding}

Transform coding is a widely-used technique for signal compression,
able to capture the transient nature of signals. Among the various
transforms, the \gls{dct} \cite{de1997data,qing_compression_2011},
the Fourier transform \cite{dapper_high_2015,Kapisch_Spectra_Variation_Based_Signal_Compressio2022},
and the \gls{dwt} \cite{littler_wavelets_1999,chung_variable_1999,hsieh2003disturbance,wu2003data,shang2003efficiency,yuan2006power,ning_wavelet-based_2011,Prathibha_Dual_tree_complex_wavelet2016}
have been mainly exploited. Other transforms include the Slantlet
transform \cite{panda2002data}, spline wavelets \cite{dash2003power,meher_integrated_2004},
wavelet packet-enhanced transforms \cite{huang2004application,Lorio_Analysis_of_data_compression},
the Hartley transform \cite{de1997data,nascimento2023hartley}, filter
banks \cite{wang_adaptive_2021}, or trainable transforms \cite{balle_end--end_2017,balle_variational_2018,lee_context-adaptive_2019,kong2023mixture,aggarwal2023neural}
adapted for electrical signal compression \cite{MMC,MMC_DCC}.

After the signal transform, only the most significant coefficients
are kept \cite{de1997data,littler_wavelets_1999,hamid_wavelet-based_2002,hsieh2003disturbance}.
Their number is determined by a rate or distortion constraint. The
position (index) and amplitude of the coefficients have to be transmitted
to the decoder for signal reconstruction. In \cite{de1997data}, significant
coefficients are iteratively selected until a distortion criterion
is met. The positions of these coefficients are encoded via run-length
coding, while their amplitudes are mapped to a codebook optimized
using the Linde--Buzo--Gray algorithm \cite{linde1980algorithm}.
In \cite{littler_wavelets_1999,hsieh2003disturbance}, only the most
significant wavelet coefficients are retained and entropy-coded via
\gls{lzw} in \cite{littler_wavelets_1999} or Huffman coding in \cite{hsieh2003disturbance}.
In \cite{hamid_wavelet-based_2002}, the most significant coefficients
are determined using the \gls{mdl} criterion \cite{rissanen1978modeling},
which allows the selection of significant coefficients by minimizing
a rate-distortion compromise. The amplitudes and positions are then
encoded using \gls{lzw} or Huffman coding. These approaches thus
easily achieve rate or distortion constraints within a margin corresponding
to the rate required to represent one significant coefficient. In
\cite{shang2003efficiency,ning_wavelet-based_2011,Prathibha_Dual_tree_complex_wavelet2016,khan_weighted_2021},
a threshold (to zero out insignificant coefficients) and the QP are
adjusted to meet a rate or distortion constraint. 

In \cite{wu2003data,yuan2006power}, vector quantization is employed.
The quantization cells are trained using a set of representative test
signals. The coefficients are then mapped to a pre-trained codebook.
The rate or distortion is adjusted by controlling the codebook size
or selecting different quantization levels for the coefficients. More
recently, in \cite{Kapisch_Spectra_Variation_Based_Signal_Compressio2022},
significant coefficients are selected based on a threshold derived
from the minimum, maximum, and mean values of the transformed coefficients,
and the retained coefficients, along with their positions, are encoded
using a lossless method like \gls{lzw} coding.

To optimize the quantization of transformed coefficients, \cite{dapper_high_2015,cormane_spectral_2016,nascimento_improved_2020,nascimento2023hartley}
exploit linear \cite{cormane_spectral_2016} or polynomial \cite{dapper_high_2015}
models approximating the envelope of the spectral coefficients to
capture the global energy distribution across frequencies. These models
yield a prediction of the amplitude of coefficients in different bands,
facilitating adaptive quantization. In \cite{cormane_spectral_2016},
a non-parametric model is considered based on spectral estimation
from a set of dictionaries whose components represent the spectral
signature of power system disturbances. Dynamic spectral prediction
models are introduced in \cite{nascimento_improved_2020,nascimento2023hartley},
based either on the variance of the coefficients or on their maximum
value.

Inspired by image compression, scalable compression schemes, such
as the \gls{ezw} bit-plane coding method, have been adapted to one-dimensional
signals in \cite{chung_variable_1999,khan_embedded-zerotree-wavelet-based_2015}.
In \cite{hoang_new_2009} as in \cite{MMC,MMC_DCC}, \gls{dct} is
followed by bit-plane coding. Variational autoencoders have been applied
in \cite{MMC,MMC_DCC}, trained on disturbance signals \cite{Database_RTE}
to optimize the end-to-end compression pipeline. Nevertheless, the rate-distortion
compromise is adjusted during training and cannot be perfectly met
for specific signals.

\subsubsection{Dictionary coding}

Dictionary coding approaches decompose the signal to compress into
a linear combination of vectors from a predefined dictionary of signals.
For example, in \cite{sabarimalai_manikandan_simultaneous_2015,he_high_2019},
the dictionaries consist of cosines, sines, and Dirac functions which,
compared to \gls{dct} or \gls{dwt}, yield a better representation
of signals with localized defects. The dictionary consists of damped
sinusoids, cosines, and Gabor functions in \cite{de_oliveira_artificial_2018}
and of cosines and wavelets in \cite{Ruiz_electrical_2020}. The combination
of these functions provides higher levels of sparsity compared to
\cite{lovisolo_modeling_2007,sabarimalai_manikandan_simultaneous_2015},
at the price of an increased computational complexity.

Matching pursuit algorithms \cite{mallat1993matching} are commonly
used to select the combination of dictionary elements representing
the signal. The encoding process stops once the required number of
selected elements is reached, making it easier to meet either rate
or distortion constraints. In \cite{de_oliveira_artificial_2018}
artificial neural network select the best sub-dictionary used at each
matching pursuit iteration.

\subsection{Rate and distortion optimization in two-stage schemes\label{subsec:More-Complex-Approach}}

In two-stage compression approaches such as \cite{ribeiro_novel_2007,zhang_high_2011,tse_real-time_2012},
considering a total bit budget, the first stage performs parametric
coding of the main signal components, while the second stage encodes
the residual using a transform-based method. One of the difficulties
is to determine the bit allocation between stages to maximize the
reconstruction quality. In \cite{ribeiro_novel_2007,zhang_high_2011,tse_real-time_2012},
a fixed number of bits is used in the first stage to represent sinusoidal
components, producing a coarse reconstruction. The residual is then
encoded using a non-parametric approach to reach the target distortion.
In \cite{hao2023compression}, the signal is decomposed into a sum
of damped sinusoids until a first distortion target is reached. The
reconstruction residual from the first stage is encoded using a DWT
followed by entropy coding until the final distortion target is attained.
Nevertheless, imposing a fixed rate or distortion target in the first
stage is not necessarily optimal: the efficiency of the first stage
is often dependent on the signal characteristics. In \cite{MMC},
the rate allocation between the two stages is optimized to get the
minimum distortion. Nevertheless, this approach is computationally
demanding, since for each choice of the rate allocation in the first
stage, a different residual is obtained with a different rate-distortion
characteristic. Moreover, no distortion target satisfaction is considered.

\section{Problem formulation}

\label{sec:Overview MMC}

This section briefly recalls the two-stage multiple-model electrical
signal compression approach introduced in \cite{MMC}. Then, the problem
of minimizing the total bit budget under a target distortion constraint
is formulated. This problem involves the joint selection of (\emph{i})
the model in the first stage, (\emph{ii}) the bit budget to encode
its estimated parameters, (\emph{iii}) the bit allocation among the
components of the parameter vector, (\emph{iv}) the residual compression
method, and (\emph{v}) the bit budget in the second stage.

\begin{figure}[htpb]
	\centering{}{\includegraphics[width=\columnwidth]{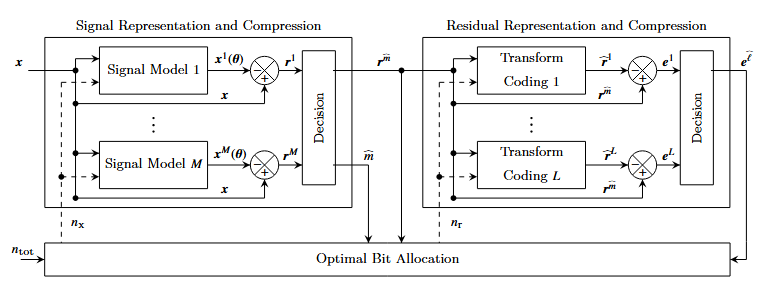}\caption{Diagram of the two-stage multiple-model electrical signal compression
			approach introduced in \cite{MMC}.}
		\label{fig:MMC}}
\end{figure}

\subsection{Two-stage multiple-model electrical signal compression approach}
\label{Ssec:MMC}

Consider an alternating single-phase signal $x\left(t\right)$ acquired
within an electrical substation at a sampling period $T_{\text{s}}=1/f_{\text{s}}$.
The sample of $x\left(t\right)$ at time $kT_{\text{s}}$, $k\in\mathbb{Z}$,
is denoted $x_{k}$. Vectors of $N$ consecutive samples $\boldsymbol{x}_{i}=\left(x_{iN+1},\dots,x_{\left(i+1\right)N}\right)^{T}$,
$i\in\mathbb{Z}$, are compressed immediately after their acquisition.
The reconstructed vector is denoted $\widehat{\boldsymbol{x}}_{i}=\left(\widehat{x}_{iN+1},\dots,\widehat{x}_{\left(i+1\right)N}\right)^{T}$,
$i\in\mathbb{Z}$. The aim is to design a coding scheme that minimizes
the \gls{mse} between $\boldsymbol{x}_{i}$ and $\widehat{\boldsymbol{x}}_{i}$.
Without loss of generality, we consider the compression of $\boldsymbol{x}_{0}=\left(x_{1},\dots,x_{N}\right)^{T}$.
Moreover, to further lighten notations, when $\boldsymbol{x}_{0}$
is processed independently of the previous vectors, the subscript
$0$ is omitted.

The compression of $\boldsymbol{x}$ is performed in two stages, presented
in Figure~\ref{fig:MMC}. In the first stage, $M$ parametric models
with output $\boldsymbol{x}^{m}\left(\boldsymbol{\theta}\right)$,
$m\in\mathcal{M}=\left\{ 1,\dots,M\right\} $, are put in competition.
The vector $\boldsymbol{\theta}\in\mathbb{R}^{K^{m}}$ of $K^{m}$
parameters of the $m$-th model is assumed to follow an \textit{a
priori} distribution $p^{m}\left(\boldsymbol{\theta}\right)$, specified
in the model. For the $m$-th model, the estimate of $\boldsymbol{\theta}$
minimizing the norm of the estimation residual $\boldsymbol{r}^{m}\left(\boldsymbol{\theta}\right)=\boldsymbol{x}-\boldsymbol{x}^{m}\left(\boldsymbol{\theta}\right)$
is
\begin{equation}
\widehat{\boldsymbol{\theta}}=\arg\min_{\boldsymbol{\theta}}\frac{1}{N}\left\Vert \boldsymbol{r}^{m}\left(\boldsymbol{\theta}\right)\right\Vert ^{2}.\label{eq:estimation}
\end{equation}

The estimate $\widehat{\boldsymbol{\theta}}$ is quantized and represented
using a budget of $n_{\text{x}}$ bits to get $\widehat{\boldsymbol{\theta}}_{\text{q}}\left(n_{\text{x}}\right)$.
The dependency of quantized vectors in the bit budget is often omitted
in what follows. The selected model $\widehat{m}$ is the one minimizing
the norm of the estimation residual $\boldsymbol{r}^{m}\left(\widehat{\boldsymbol{\theta}}_{\text{q}}\right)$,
evaluated considering the quantized version of the estimated parameter
vector. A coarse representation $\boldsymbol{x}^{\widehat{m}}\left(\widehat{\boldsymbol{\theta}}_{\text{q}}\right)$
of $\boldsymbol{x}$ is then obtained at the output at the first stage,
as well as $\boldsymbol{r}^{\widehat{m}}\left(\widehat{\boldsymbol{\theta}}_{\text{q}}\right)$,
to be processed by the second stage.

Many models can be considered in the first stage, such as the sinusoidal
model
\begin{equation}
x_{n}\left(\boldsymbol{\theta}\right)=a\cos\left(2\pi fnT_{\text{s}}+\phi\right),\quad n=1,\dots,N,\label{eq:sin}
\end{equation}
with $\boldsymbol{\theta}=\left(a,f,\phi\right)^{T}$ or linear combinations
of $K$ Tchebychev polynomials of the first kind $\mathcal{T}_{k}$
of degree $k$ \cite{Kamps_chebyshev_polynomials_1989}, with output
\begin{equation}
x_{n}\left(\boldsymbol{\theta}\right)=\sum_{k=0}^{K-1}\theta_{k+1}\mathcal{T}_{k}\left(2\frac{n}{N}-1\right),\quad n=1,\dots,N,\label{eq:poly}
\end{equation}
and vector of parameters $\boldsymbol{\theta}=\left(\theta_{1},\dots,\theta_{K}\right)^{T}$.
Additional sample predictive and parameter predictive models have
been considered in \cite{MMC_DCC,MMC} to represent the samples of
$\boldsymbol{x}_{i}$ using previously encoded samples or previously
considered models. In the sample predictive model, a model of $\boldsymbol{x}_{i}$
is built using the reconstructed samples $\widehat{x}_{iN}$, $\widehat{x}_{iN-1}\dots$
considering a linear predictor of order $K$, offset $\eta\geqslant0$,
and gains $\theta_{1},\dots,\theta_{K}$ as 
\begin{equation}
x_{n}\left(\boldsymbol{\theta}\right)=\sum_{k=1}^{K}\theta_{k}\widehat{x}_{\left(i-1\right)N+n-\eta-k+1},\quad n=1,\dots,N,\label{eq:samples_pred}
\end{equation}
and $\boldsymbol{\theta}=\left(\theta_{1},\dots,\theta_{K}\right)^{T}$.
In the parameter predictive model, to represent $\boldsymbol{x}_{i}$,
the model of index $\widehat{m}_{i-1}$ and quantized parameter vector
$\widehat{\boldsymbol{\theta}}_{i-1,\text{q}}$ is reused and a differential
coding of the vector of parameters is performed by evaluating
\begin{equation}
\widehat{\boldsymbol{\delta}}=\arg\min_{\boldsymbol{\delta}}\frac{1}{N}\left\Vert \boldsymbol{x}_{i}-\boldsymbol{x}^{\widehat{m}_{i-1}}\left(\widehat{\boldsymbol{\theta}}_{i-1,\text{q}}+\boldsymbol{\delta}\right)\right\Vert ^{2}.\label{eq:para_pred}
\end{equation}
Then $\widehat{\boldsymbol{\delta}}$ is quantized and represented
using a budget of $n_{\text{x}}$ bits to get $\widehat{\boldsymbol{\delta}}_{\text{q}}$.
The model output is obtained using \eqref{eq:sin}, \eqref{eq:poly},
or \eqref{eq:samples_pred} with $\widehat{\boldsymbol{\theta}}_{i,\text{q}}=\widehat{\boldsymbol{\theta}}_{i-1,\text{q}}+\widehat{\boldsymbol{\delta}}_{\text{q}}$.
Finally, the option of bypassing the first stage is considered when
no parametric model can adequately represent the original signal.
In this case, a null model with output 
\begin{equation}
x_{n}\left(\boldsymbol{\theta}\right)=0,\quad n=1,\dots,N,\label{eq:none_model}
\end{equation}
is employed, with $\boldsymbol{\theta}=\emptyset$.

In the second stage, $L$ residual compression methods are compared,
each of which providing an estimate $\widehat{\boldsymbol{r}}^{\ell}$
of $\boldsymbol{r}^{\widehat{m}}\left(\widehat{\boldsymbol{\theta}}_{\text{q}}\right)$.
The method $\widehat{\ell}$ that yields the smallest norm of the
residual $\boldsymbol{e}^{\ell}=\boldsymbol{r}^{\widehat{m}}\left(\widehat{\boldsymbol{\theta}}_{\text{q}}\right)-\widehat{\boldsymbol{r}}^{\ell}$
for a given bit budget $n_{\text{r}}$ is selected. In addition to
\gls{dct} and \gls{dwt}, trainable transforms based on \gls{vae}
\cite{balle_variational_2018,alves2021reduced,kong2023mixture} are
also considered. The second stage may also be bypassed when the first
stage represents efficiently enough the signal $\boldsymbol{x}$ (when
the distortion target is met).

Among the $n_{\text{tot}}$ bits available to represent $\boldsymbol{x}$,
$n_{\text{x}}$ are used for the quantized model parameter vector,
$n_{\text{r}}$ for the residual, and $n_{\text{h}}$ for all additional
(header) information required to reconstruct $\boldsymbol{x}$. The
allocation of $n_{\text{tot}}$ among $n_{\text{h}}$, $n_{\text{x}}$,
and $n_{\text{r}}$ has thus to be optimized.

The dynamic range of signals and residuals may vary significantly.
To address this issue, $\boldsymbol{x}$ (before the first stage)
and $\boldsymbol{r}^{\widehat{m}}\left(\widehat{\boldsymbol{\theta}}_{\text{q}}\right)$
(before the second stage) are multiplied by $2^{k_{\text{x}}}$ and
$2^{k_{\text{r}}}$ respectively, where the scaling coefficients $k_{\text{x}}\in\mathbb{Z}$
and $k_{\text{r}}\in\mathbb{Z}$ are such that $\max\left(2^{k_{\text{x}}}\left\vert \boldsymbol{x}\right\vert \right)\in\left[0.5,1\right[$
and $\max\left(2^{k_{\text{r}}}\left\vert \boldsymbol{r}^{\widehat{m}}\left(\widehat{\boldsymbol{\theta}}_{\text{q}}\right)\right\vert \right)\in\left[0.5,1\right[$.
The parameters $k_{\text{x}}$ and $k_{\text{r}}$ are stored in the
header, with a fixed-length representation, using part of the $n_{\text{h}}$
bits.

\subsection{Optimization of the compression parameters under a distortion constraint\label{sec:Problem formulation}}

For a given vector $\boldsymbol{x}$, consider the first-stage model
output $\boldsymbol{x}^{m}\left(\widehat{\boldsymbol{\theta}}_{\text{q}}\left(n_{\text{x}}\right)\right)$
where $\widehat{\boldsymbol{\theta}}$ has been represented with $n_{\text{x}}$
bits and the second-stage reconstructed residual $\widehat{\boldsymbol{r}}^{\ell}\left(n_{\text{r}}\right)$
with a bit budget $n_{\text{r}}$. To recover an estimate of the original
vector from the output of both stages, $n_{\text{h}}$ bits are required
to represent $n_{\text{x}}$ and $n_{\text{r}}$ themselves, the scaling
factors $k_{\text{x}}$ and $k_{\text{r}}$, and the indices of the
best model $\widehat{m}$ and the best compression method $\widehat{\ell}$.
We assume that $m$, $\ell$, $k_{\text{x}}$, $k_{\text{r}}$, $n_{\text{x}}$,
and $n_{\text{r}}$ are represented with a fixed-length code in the
header and that $n_{\text{h}}$ is therefore constant. 

Our aim in what follows is, considering some target distortion constraint
$D_{\text{max}}$, to determine the optimal values of $m$, $\ell$,
$n_{\text{x}}$, and $n_{\text{r}}$, which are the solution of the
following constrained optimization problem 
\begin{align}
\widehat{m},\widehat{\ell},\widehat{n}_{\text{x}},\widehat{n}_{\text{r}} & =\arg\min_{m,\ell,n_{\text{x}},n_{\text{r}}}n_{\text{h}}+n_{\text{x}}+n_{\text{r}}\label{eq:optimization_problem}\\
 & \text{s.t. }D^{m,\ell}\left(n_{\text{x}},n_{\text{r}}\right)\leqslant D_{\text{max}}.\nonumber 
\end{align}
where
\begin{equation}
D^{m,\ell}\left(n_{\text{x}},n_{\text{r}}\right)=\frac{1}{N}\left\Vert \boldsymbol{x}-\boldsymbol{x}^{m}\left(\widehat{\boldsymbol{\theta}}_{\text{q}}\left(n_{\text{x}}\right)\right)-\widehat{\boldsymbol{r}}^{\ell}\left(n_{\text{r}}\right)\right\Vert ^{2}\label{eq:Distorsion}
\end{equation}
is the distortion obtained at the output of the second stage, when
the $m$-th model is used with parameter vector $\widehat{\boldsymbol{\theta}}$
quantized with $n_{\text{x}}$ bits and when the $\ell$-th residual
compression method is used with $n_{\text{r}}$ bits.

\section{Proposed Solutions\label{sec:Proposed solutions}}

To address the optimization problem in \eqref{eq:optimization_problem},
Section~\ref{subsec:Optimal representation of the model parameters}
first describes, for a given value of $n_{\text{x}}$, the allocation
of these bits across components of the quantized model parameter vector.
Since the value of $n_{\text{x}}$, solution of \eqref{eq:optimization_problem},
is not known \emph{a priori}, Section~\ref{subsec:interval} determines
a search interval for candidate values of $n_{\text{x}}$ so as to
reduce the number of cases to consider. Section~\ref{subsec:Exhaustive Search}
details an \gls{es} method to identify the allocation of
$n_{\text{x}}$ and $n_{\text{r}}$, achieving an optimal solution
for \eqref{eq:optimization_problem}. Finally, Section~\ref{sec:Golden Section Search}
describes a reduced-complexity approach based on \gls{gss}.

\subsection{Representation of the Quantized Model Parameters\label{subsec:Optimal representation of the model parameters}}

In the first compression stage, consider the $m$-th model and a budget
of $n_{\text{x}}$ bits to represent the vector of parameters $\widehat{\boldsymbol{\theta}}$.
In this paper, a scalar quantization of each component of $\widehat{\boldsymbol{\theta}}$
is performed to get $\widehat{\boldsymbol{\theta}}_{\text{q}}\left(n_{1},\dots,n_{K}\right)$,
where $n_{k}$ is the number of bits used to represent the $k$-th
parameter $\widehat{\theta}_{k}$. The quantization of $\widehat{\boldsymbol{\theta}}$
introduces an error in the model output expressed as 
\begin{equation}
\boldsymbol{\varepsilon}_{\mathrm{q}}^{m}\left(\widehat{\boldsymbol{\theta}},\left(n_{1},\dots,n_{K}\right)\right)=\boldsymbol{x}^{m}\left(\widehat{\boldsymbol{\theta}}\right)-\boldsymbol{x}^{m}\left(\widehat{\boldsymbol{\theta}}_{\text{q}}\left(n_{1},\dots,n_{K}\right)\right).\label{eq:eps_q}
\end{equation}
Our aim is to determine the allocation of bits $n_{k}$, $k=1,\dots,K$,
that minimizes the expected norm of the error introduced by the quantization
of $\widehat{\boldsymbol{\theta}}$ on the model output 
\begin{align}
\widehat{n}_{1},\dots,\widehat{n}_{K}= & \arg\min_{n_{1},\dots,n_{K}}\mathbb{E}_{\widehat{\boldsymbol{\theta}}}\left[\frac{1}{N}\left\Vert \boldsymbol{x}^{m}\left(\widehat{\boldsymbol{\theta}}\right)-\boldsymbol{x}^{m}\left(\widehat{\boldsymbol{\theta}}_{\text{q}}\left(n_{1},\dots,n_{K}\right)\right)\right\Vert ^{2}\right]\label{eq:allocation_rate_parameters}\\
 & \text{s.t.}~\sum_{k=1}^{K}n_{k}=n_{\text{x}},\nonumber 
\end{align}
where the expectations is evaluated with respect to the \textit{a
priori} distribution of $\widehat{\boldsymbol{\theta}}$. The aim
is to get values $\widehat{n}_{k}$ for $k=1,\dots,K$ that are independent
of the signal samples to be encoded and are only characteristics of
the model $m$ for a given $n_{\text{x}}$. This avoids transmitting
$\widehat{n}_{k}$, for $k=1,\dots,K$, provided that an offline optimization
has been done for a sufficiently large variety of values of $n_{\text{x}}$.

Proposition~\ref{prop:model_eq} provides an approximate expression
of the expected value of the norm of the error induced by the quantization
of $\widehat{\boldsymbol{\theta}}$.
\begin{prop}
\label{prop:model_eq} Assuming that the components of the quantization
error $\boldsymbol{\varepsilon}=\widehat{\boldsymbol{\theta}}-\widehat{\boldsymbol{\theta}}_{\mathrm{q}}$
of the parameter vector $\widehat{\boldsymbol{\theta}}$ are zero-mean
and uncorrelated and that they are uncorrelated with $\frac{\partial x_{n}^{m}\left(\boldsymbol{\theta}\right)}{\partial\theta_{k}}$
for all $k=1,\dots,K$ and $n=1,\dots,N$, one has 
\begin{equation}
\mathbb{E}_{\widehat{\boldsymbol{\theta}}}\left[\frac{1}{N}\left\Vert \boldsymbol{x}^{m}\left(\widehat{\boldsymbol{\theta}}\right)-\boldsymbol{x}^{m}\left(\widehat{\boldsymbol{\theta}}_{\mathrm{q}}\left(n_{1},\dots,n_{k}\right)\right)\right\Vert ^{2}\right]=\sum_{k=1}^{K}h_{k}\mathbb{E}_{\widehat{\theta}_{k}}\left[\varepsilon_{k}^{2}\right],\label{eq:model_e_q}
\end{equation}
where $\varepsilon_{k}$ is the $k$-th component of $\boldsymbol{\varepsilon}$
and
\begin{equation}
h_{k}=\mathbb{E}_{\widehat{\boldsymbol{\theta}}}\left[\frac{1}{N}\left\Vert \frac{\partial\boldsymbol{x}^{m}\left(\boldsymbol{\theta}\right)}{\partial\theta_{k}}\right\Vert ^{2}\right].\label{eq:model_e_q_pk}
\end{equation}
\end{prop}
\begin{proof}
Proposition~\ref{prop:model_eq} is a special case of Proposition~\ref{prop:model_eq2}, which proof is in Appendix~\ref{subsec:Proof of proposition model eq2}.
\end{proof}
The components $h_{k}$ introduced in \eqref{eq:model_e_q_pk} are
the expected values of the squared norm of the sensitivity of $\boldsymbol{x}^{m}\left(\boldsymbol{\theta}\right)$
with respect to $\theta_{k}$ and depend on the chosen model. It has
been shown in \cite{MMC} that for a sinusoidal model, $h_{1}=1/2$,
$h_{2}=\mathbb{E}\left[a^{2}\right]\left(2\pi NT_{\text{s}}\right)^{2}/6$,
and $h_{3}=\mathbb{E}\left[a^{2}\right]/2$. For a Tchebychev polynomial
model of degree $k$, $h_{k}=\frac{1}{N}\sum_{n=1}^{N}\mathcal{T}_{k}^{2}\left(2n/N-1\right)$,
$k=0,\dots,K-1$. For a sample predictive model of order $k$ and
offset $\eta$, $h_{k}=\frac{1}{N}\sum_{n=1}^{N}\widehat{x}_{\left(i-1\right)N+n-\eta-k+1}^{2}$,
$k=1,\dots,K$. For parameter predictive model, the entries of $\widehat{\boldsymbol{\delta}}$
are scalar quantized to get $\widehat{\boldsymbol{\delta}}_{\text{q}}=\widehat{\boldsymbol{\delta}}+\boldsymbol{\varepsilon}$,
where $\boldsymbol{\varepsilon}$ is the quantization error vector.
Consequently, the optimal bit allocation for the quantization can
be done considering $\widehat{\boldsymbol{\delta}}$ instead of $\widehat{\boldsymbol{\theta}}$,
whatever the model.

Assuming a smooth distribution $p\left(\widehat{\boldsymbol{\theta}}\right)$,
at high rate one gets $\mathbb{E}_{\widehat{\boldsymbol{\theta}}}\left[\varepsilon_{k}^{2}\right]=c_{k}^{2}2^{-2n_{k}}$
\cite{panter1951quantization}, with 
\begin{equation}
c_{k}^{2}=\frac{1}{12}\left[\int_{\mathbb{R}}\sqrt[3]{p\left(\widehat{\theta}_{k}\right)}d\widehat{\theta}_{k}\right]^{3}.
\end{equation}
For example, if $\widehat{\theta}_{k}$ is uniformly distributed over
an interval of width $w_{k}$, then $c_{k}^{2}=\frac{w_{k}^{2}}{12}$.
If $\widehat{\theta}_{k}$ follows a Gaussian distribution of variance
$\sigma_{k}^{2}$, then $c_{k}^{2}=\frac{\sqrt{3}\pi}{2}\sigma_{k}^{2}$.
\begin{prop}
\label{prop:nk} Consider a model with $K$ parameters. Assume that
the $k$-th parameter is scalar quantized with $n_{k}$ bits and that
$\mathbb{E}_{\widehat{\theta}_{k}}\left[\left(\widehat{\theta}_{k}-\widehat{\theta}_{\mathrm{q},k}\right)^{2}\right]=c_{k}^{2}2^{-2n_{k}}$
for some $c_{k}$. Assume without loss of generality that $p_{1}c_{1}^{2}\geqslant\dots\geqslant h_{K}c_{K}^{2}$.
Then, the solution of \eqref{eq:allocation_rate_parameters} is 
\begin{align}
\widehat{n}_{k} & =\frac{n_{\mathrm{x}}}{K'}+\frac{1}{2}\log_{2}\frac{h_{k}c_{k}^{2}}{\left(\prod_{j=1}^{K'}h_{j}c_{j}^{2}\right)^{1/K'}},\quad k=1,\dots,K',\label{eq:nk}\\
\widehat{n}_{k} & =0,\quad k=K'+1,\dots,K,
\end{align}
where $K'\leqslant K$ is the largest integer such that $\widehat{n}_{k}\geqslant0,\quad k=1,\dots,K'$.
\end{prop}
\begin{proof}
The proof is similar to that for the bit allocation for independent
Gaussian sources minimizing the total quadratic distortion, where
the variance of the sources is replaced by $h_{k}c_{k}^{2}$, see
\cite[Chap.13.5]{sayood_introduction_2017}.
\end{proof}
The expression \eqref{eq:nk} takes into account the sensitivity of
the model with respect to its parameters. More rate is allocated to
parameters for which the term $h_{k}c_{k}^{2}$ is larger than the
geometric mean of the $K'$ largest terms.

In what follows, $\boldsymbol{\varepsilon}_{\mathrm{q}}^{m}\left(\widehat{\boldsymbol{\theta}},n_{\mathrm{x}}\right)$
represents the vector $\boldsymbol{\varepsilon}_{\mathrm{q}}^{m}\left(\widehat{\boldsymbol{\theta}},\left(\widehat{n}_{1},\dots,\widehat{n}_{K}\right)\right)$,
with $\widehat{n}_{1},\dots,\widehat{n}_{K}$ obtained from Proposition~\ref{prop:nk}.

\subsection{Search Interval for $n_{\text{x}}$}

\label{subsec:interval}

An upper bound $n_{\max}$ for the number of bits $n_{\text{x}}$
to use in the first stage can be obtained by evaluating the rate that
would be required by the second stage to meet the target distortion
when skipping the first stage. In such case, $\boldsymbol{x}$ is
directly represented by $\widehat{\boldsymbol{r}}^{\ell}(n)$ and
for the $\ell$-th residual compression method, one has
\begin{equation}
n_{\text{max}}^{\ell}=\min n\quad\text{s.t.}\quad\frac{1}{N}\left\Vert \boldsymbol{x}-\widehat{\boldsymbol{r}}^{\ell}(n)\right\Vert ^{2}\leqslant D_{\text{max}},\label{eq:nmax}
\end{equation}
\begin{equation}
n_{\max}=\min_{\ell}n_{\text{max}}^{\ell}\label{eq:nmaxreal}
\end{equation}
and
\begin{equation}
\underline{\ell}=\arg\min_{\ell}n_{\text{max}}^{\ell}.\label{eq:argnmax}
\end{equation}

Now, consider the model $m$ in the first stage. If the quantization
error of the estimated parameter vector $\widehat{\boldsymbol{\theta}}$
is neglected, the rate required by the second stage to represent the
residual $\boldsymbol{x}-\boldsymbol{x}^{m}\left(\widehat{\boldsymbol{\theta}}\right)$
with the target distortion would be
\begin{equation}
n_{\text{min}}^{m}=\min_{\ell}n_{\text{min}}^{m,\ell},
\end{equation}
where 
\begin{equation}
n_{\text{min}}^{m,\ell}=\min n\quad\text{s.t.}\quad\frac{1}{N}\left\Vert \boldsymbol{x}-\boldsymbol{x}^{m}(\widehat{\boldsymbol{\theta}})-\widehat{\boldsymbol{r}}^{\ell}(n)\right\Vert ^{2}\leqslant D_{\text{max}}\label{eq:nmin}
\end{equation}
is the rate using the $\ell$-th compression method. In practice,
the rate required by the second stage is in general larger than $n_{\text{min}}^{m,\ell}$,
to compensate the impact of the quantization error, since, according
to \eqref{eq:estimation}, one has 
\begin{equation}
\frac{1}{N}\left\Vert \boldsymbol{x}-\boldsymbol{x}^{m}(\widehat{\boldsymbol{\theta}}_{\text{q}}\left(n_{\text{x}}\right))\right\Vert ^{2}\geqslant\frac{1}{N}\left\Vert \boldsymbol{x}-\boldsymbol{x}^{m}(\widehat{\boldsymbol{\theta}})\right\Vert ^{2}.
\end{equation}
Moreover, considering $n_{\text{x}}>n_{\max}-n_{\text{min}}^{m}$
is useless as $n_{\text{x}}+n_{\text{min}}^{m}>n_{\text{max}}$ bits
would be required by the two stages, which is more than the amount
$n_{\max}$ of bits required considering only the second stage. Therefore,
for a given model $m$, the search space for the number of bits $n_{\text{x}}$
for the first stage can be reduced to the interval $\mathcal{N}_{\text{x}}^{m}=\left[0,\dots,n_{\text{max}}-n_{\text{min}}^{m}\right]$.

\begin{figure}[!h]
\centering\includegraphics[width=0.8\columnwidth]{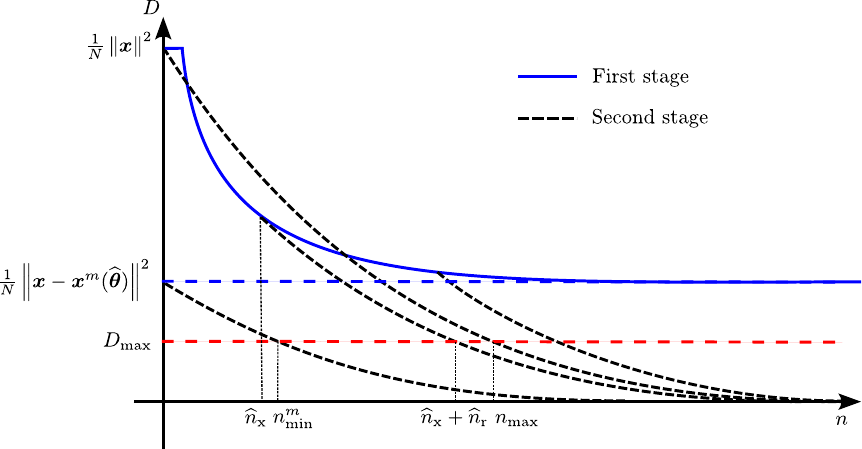} \caption{Evolution of the distortion as a function of $n_{\text{x}}$ for the
first stage and $n_{\text{x}}+n_{\text{r}}$ for the first and the
second stages. The blue curve shows the distortion of the first stage
as a function of $n_{\text{x}}$. The dotted black curves represent
the residual distortion of the output of the second stage for different
values of $n_{\text{x}}$ and as a function of $n_{\text{x}}+n_{\text{r}}$.}
\label{fig:search interval}
\end{figure}

Figure~\ref{fig:search interval} illustrates the evolution of distortion
as a function of $n_{\text{x}}$ for the first stage (blue curve)
and of $n_{\text{x}}+n_{\text{r}}$ for the combined first and second
stages (black dashed curves). The distortion at the output of the
first stage when the $m$-th model is considered reaches the minimum
$\frac{1}{N}\left\Vert \boldsymbol{x}-\boldsymbol{x}^{m}\left(\widehat{\boldsymbol{\theta}}\right)\right\Vert ^{2}$
corresponding to the absence of quantization of the estimated parameter
vector $\widehat{\boldsymbol{\theta}}$. The dotted black curves represent
the residual distortion at the output of the second stage for different
values of $n_{\text{x}}$ as a function of $n_{\text{x}}+n_{\text{r}}$.
The black dashed curve starting at $D=\frac{1}{N}\left\Vert \boldsymbol{x}\right\Vert ^{2}$
represents the case where only the second stage is used. The point
at which this curve intersects the red dashed line representing $D_{\text{max}}$
yields $n_{\text{max}}$. The dotted black curve that begins at $\frac{1}{N}\left\Vert \boldsymbol{x}-\boldsymbol{x}^{m}\left(\widehat{\boldsymbol{\theta}}\right)\right\Vert ^{2}$
represents the residual distortion of the output of the second stage
when the quantization error at the first stage is neglected ($\widehat{\boldsymbol{\theta}}_{\text{q}}\left(n_{\text{x}}\right)=\widehat{\boldsymbol{\theta}}$).
The minimum rate $n_{\text{min}}^{m}$ is determined at the intersection
of this curve with $D_{\text{max}}$. The two remaining black curves
illustrate other scenarios for different values of $n_{\text{x}}$,
demonstrating that the total number of bits required to satisfy the
distortion constraint can vary depending on the choice of $n_{\text{x}}$.

\subsection{Exhaustive Search\label{subsec:Exhaustive Search}}

In this section, the optimization problem \eqref{eq:optimization_problem}
is solved via an \gls{es}, as detailed in Algorithm~\ref{alg:optimal_solution}.

First, the upper bound $n_{\max}$ is determined using \eqref{eq:nmax}
(line 1). For each model $m\in\mathcal{M}$ (line 2), $\widehat{\boldsymbol{\theta}}$
is estimated using \eqref{eq:estimation} (line 3), and $n_{\text{min}}^{m}$
is evaluated using \eqref{eq:nmin} (line 4). The set of candidate
values $\mathcal{N}_{\text{x}}^{m}$ for $n_{\text{x}}$ is then obtained
(line 5).

For each $n_{\text{x}}\in\mathcal{N}_{\text{x}}^{m}$, $\widehat{\boldsymbol{\theta}}$
is quantized to get $\widehat{\boldsymbol{\theta}}_{\mathrm{q}}\left(n_{\text{x}}\right)$
using the bit allocation from Proposition~\ref{prop:nk} (line 7).
Then, for each residual compression method $\ell\in\mathcal{L}$,
the distortion 
\begin{equation}
D^{m,\ell}\left(n_{\text{x}},n_{\text{r}}\right)=\frac{1}{N}\left\Vert \boldsymbol{x}-\boldsymbol{x}^{m}\left(\widehat{\boldsymbol{\theta}}_{\mathrm{q}}\left(n_{\text{x}}\right)\right)-\widehat{\boldsymbol{r}}^{\ell}\left(n_{\text{r}}\right)\right\Vert ^{2}\label{eq:Distortion}
\end{equation}
is evaluated for increasing values of $n_{\text{r}}$ (line 8) until
the distortion constraint $D_{\text{max}}$ is satisfied for $\underline{n}_{\text{r}}^{\ell}$.
Then, $n_{\text{r}}\left(n_{\text{x}}\right)$, the smallest value
of $\underline{n}_{\text{r}}^{\ell}$, $\ell\in\mathcal{L}$ and the
corresponding index of the second-stage compression approach $\underline{\ell}$
are stored (line 11). If, for $n_{\text{x}}$, the model $m$ is such
that $n_{\text{x}}+n_{\text{r}}\left(n_{\text{x}}\right)<n_{\text{max}}$,
then $n_{\text{max}}$ and $\mathcal{N}_{\text{x}}^{m}$ are updated
(lines 13 and 14). Moreover, the values of $n_{\text{x}}$, $n_{\text{r}}$,
the corresponding model $m$, and residual compression method $\ell$
are stored as candidate output of the algorithm (line 15).

\begin{algorithm}[!h]
\caption{Optimal solution for \eqref{eq:optimization_problem} via exhaustive search}
\label{alg:optimal_solution} 
Determine $n_{\max}$ using \eqref{eq:nmax}\;

\For{$m\in\mathcal{M}$}{

Evaluate $\widehat{\boldsymbol{\theta}}$ using \eqref{eq:estimation}
\label{line:1}\;

Evaluate $n_{\text{\ensuremath{\min}}}^{m}$ using \eqref{eq:nmin}
\;

Initialize $\mathcal{N}_{\text{x}}^{m}=\left[0,n_{\text{\ensuremath{\max}}}-n_{\text{\ensuremath{\min}}}^{m}\right]$\;

\For{each $n_{\text{x}}\in\mathcal{N}_{\text{x}}^{m}$}{

Quantize $\widehat{\boldsymbol{\theta}}$ to get $\widehat{\boldsymbol{\theta}}_{\mathrm{q}}\left(n_{\text{x}}\right)$
using the bit allocation of Proposition~\ref{prop:nk}\;

\For{each $\ell\in\mathcal{L}$}{

Find $\underline{n}_{\text{r}}^{\ell}$, the smallest value of $n_{\text{r}}^{\ell}$
such that $\frac{1}{N}\left\Vert \boldsymbol{x}-\boldsymbol{x}^{m}\left(\widehat{\boldsymbol{\theta}}_{\mathrm{q}}\left(n_{\text{x}}\right)\right)-\widehat{\boldsymbol{r}}^{\ell}\left(n_{\text{r}}^{\ell}\right)\right\Vert ^{2}\leqslant D_{\text{max}}$\;
}

Evaluate $n_{\text{r}}\left(n_{\text{x}}\right)=\min_{\ell}\underline{n}_{\text{r}}^{\ell}$
and $\underline{\ell}=\arg\min_{\ell}\underline{n}_{\text{r}}^{\ell}$\;

\If{$n_{\text{x}}+n_{\text{r}}\left(n_{\text{x}}\right)<n_{\max}$}{

$n_{\max}=n_{\text{x}}+n_{\text{r}}\left(n_{\text{x}}\right)$\;

$\mathcal{N}_{\text{x}}^{m}=\left[0,n_{\text{\ensuremath{\max}}}-n_{\text{\ensuremath{\min}}}^{m}\right]$

$\widehat{n}_{\text{x}}=n_{\text{x}}$; $\widehat{n}_{\text{r}}=n_{\text{r}}\left(n_{\text{x}}\right)$;
$\widehat{m}=m$; $\widehat{\ell}=\underline{\ell}$ \tcc*{Update output} 
} } }
\end{algorithm}

If the compression approach $\ell$ is scalable, then the minimum
value $n_{\text{r}}^{\ell}$ is easily obtained at line 8 of Algorithm~\ref{alg:optimal_solution}
via a single encoding step. The value of $n_{\max}$ may decrease
from one model to another, leading to a decrease of the size of the
interval $\mathcal{N}_{\text{x}}^{m}$, as models are tested.

Regarding the complexity, for each model $m\in\mathcal{M}$, the estimate
$\widehat{\boldsymbol{\theta}}$ of the vector of parameters has to
be evaluated only once. When the model output $\boldsymbol{x}^{m}\left(\boldsymbol{\theta}\right)$
is linear in $\boldsymbol{\theta}$, an explicit expression can be
used to obtain $\widehat{\boldsymbol{\theta}}$. When it is non-linear
in $\boldsymbol{\theta}$, an iterative search has to be performed.
Then for each model, in the worst case, $\left|\mathcal{N}_{\text{x}}^{m}\right|$
values for $n_{\text{x}}$ have to be considered, where $\left|\mathcal{N}_{\text{x}}^{m}\right|$
represents the cardinal number of the set $\mathcal{N}_{\text{x}}^{m}$.
For each of these values, Proposition~\ref{prop:nk} yields the optimal
bit allocation for the quantization of $\widehat{\boldsymbol{\theta}}$
and $\boldsymbol{x}^{m}\left(\widehat{\boldsymbol{\theta}}_{\mathrm{q}}\left(n_{\text{x}}\right)\right)$
has to be evaluated as well as the residual $\boldsymbol{r}^{m}\left(\widehat{\boldsymbol{\theta}}_{\mathrm{q}}\left(n_{\text{x}}\right)\right)$.
Then, $L$ residual compression methods have to be compared. Consequently,
in a worst-case scenario, Algorithm~\ref{alg:optimal_solution} involves
performing $\sum_{m=1}^{M}L\left|\mathcal{N}_{\text{x}}^{m}\right|$
residual compressions.

\subsection{Golden Section Search\label{sec:Golden Section Search}}

For each model $m\in\mathcal{M}$ and each value of $n_{\text{x}}\in\mathcal{N}_{\text{x}}^{m}$,
Algorithm~\ref{alg:golden_section_solution} evaluates $n_{\text{r}}\left(n_{\text{x}}\right)$,
the minimum rate required in the second stage to reach the distortion
target $D_{\text{max}}$. In this section, we assume for each model
$m\in\mathcal{M}$ that $n_{\text{x}}+n_{\text{r}}(n_{\text{x}})$
is convex in $n_{\text{x}}$. Then we solve the optimization problem
in \eqref{eq:optimization_problem} using the \gls{gss}
method \cite{PressC10:2011}, see Algorithm~\ref{alg:golden_section_solution}.
The function \texttt{Eval\_nr} (lines 1--8) evaluates $n_{\text{r}}(n_{\text{x}})$
and the associated method $\underline{\ell}$ at the second stage
for given $m$, $\widehat{\boldsymbol{\theta}}$, $n_{\text{x}}$,
and $D_{\text{max}}$. This function corresponds to lines 8--11 of
Algorithm~\ref{alg:optimal_solution}. In Algorithm~\ref{alg:golden_section_solution},
for each model $m\in\mathcal{M}$ (line 10), $\widehat{\boldsymbol{\theta}}$
is evaluated (line 11), and the interval $\mathcal{N}_{\text{x}}^{m}$
is initialized (line 13). Then, the \texttt{GoldenSectionSearch} algorithm
(line 14) is applied to find the minimum of $n_{\text{x}}+n_{\text{r}}(n_{\text{x}})$
under the distortion constraint $D^{m,\ell}\left(n_{\text{x}},n_{\text{r}}\right)\leqslant D_{\text{max}}$.
This function takes as input $m$, $\widehat{\boldsymbol{\theta}}$,
$\mathcal{N}_{\text{x}}^{m}$, and $D_{\text{max}}$ and returns $n_{\text{x}},n_{\text{r}}$,
and $\underline{\ell}$ with less function evaluations than the \gls{es}. At each iteration of the \gls{gss}, the function
\texttt{Eval\_nr} is called to get, for a given value of $n_{\text{x}}$,
the minimum of $n_{\text{x}}+n_{\text{r}}(n_{\text{x}})$ under the
distortion constraint $D^{m,\ell}\left(n_{\text{x}},n_{\text{r}}\right)\leqslant D_{\text{max}}$.

If that value is less than $n_{\max}$ (line 15), then $\widehat{n}_{\text{x}}$,
$\widehat{n}_{\text{r}}$, $\widehat{m}$, and $\widehat{\ell}$ are
updated (line 16). Finally, $n_{\text{max}}$ is updated (line 17).

\begin{algorithm}[htpb]
\caption{Solution for \eqref{eq:optimization_problem} using \gls{gss}}
\label{alg:golden_section_solution} 

\SetKwFunction{Evalnr}{$f$}

\SetKwProg{Fn}{Function}{}{end}

\Fn{\texttt{\emph{Eval\_nr}}($m$,$\widehat{\boldsymbol{\theta}}$,$n_{\text{x}}$,$D_{\max}$)}{

Quantize $\widehat{\boldsymbol{\theta}}$ to obtain $\widehat{\boldsymbol{\theta}}_{\text{q}}\left(n_{\text{x}}\right)$
with bit allocation determined by Proposition~\ref{prop:nk} \;

\For{each $\ell\in\mathcal{L}$}{

Find $\underline{n}_{\text{r}}^{\ell}$, the smallest value of $n_{\text{r}}^{\ell}$
such that $\frac{1}{N}\left\Vert \boldsymbol{x}-\boldsymbol{x}^{m}\left(\widehat{\boldsymbol{\theta}}_{\mathrm{q}}\left(n_{\text{x}}\right)\right)-\widehat{\boldsymbol{r}}^{\ell}\left(n_{\text{r}}^{\ell}\right)\right\Vert ^{2}\leqslant D_{\text{max}}$\;
}

Evaluate $n_{\text{r}}\left(n_{\text{x}}\right)=\min_{\ell}\underline{n}_{\text{r}}^{\ell}$
and $\underline{\ell}=\arg\min_{\ell}\underline{n}_{\text{r}}^{\ell}$\;

\Return{$n_{\text{r}}\left(n_{\text{x}}\right)$, $\underline{\ell}$}\;}

Determine $n_{\text{max}}$ using \eqref{eq:nmax} and $\underline{\ell}$
using \eqref{eq:argnmax}\;

\For{$m\in\mathcal{M}$}{

Evaluate $\widehat{\boldsymbol{\theta}}$ using $\eqref{eq:estimation}$\;

Evaluate $n_{\text{min}}^{m}$ using $\eqref{eq:nmin}$ \; 

Initialize $\mathcal{N}_{\text{x}}^{m}=\left[0,n_{\text{max}}-n_{\text{min}}^{m}\right]$\;

$\left(n_{\text{x}},n_{\text{r}},\underline{\ell}\right)=\text{\texttt{GoldenSectionSearch}}\left(m,\widehat{\boldsymbol{\theta}},\mathcal{N}_{\text{x}}^{m},D_{\text{max}}\right)$ \tcc*{Using Eval\_nr}

\If{$n_{\text{x}}+n_{\text{r}}<n_{\text{max}}$}{

$\widehat{n}_{\text{x}}=n_{\text{x}}$, $\widehat{n}_{\text{r}}=n_{\text{r}}$,
$\widehat{m}=m$, $\widehat{\ell}=\underline{\ell}$ \tcc*{Update
output}

$n_{\text{max}}=n_{\text{x}}+n_{\text{r}}$}

}
\end{algorithm}

Regarding the complexity, for each model $m\in\mathcal{M}$, the estimate
$\widehat{\boldsymbol{\theta}}$ of the vector of parameters has to
be evaluated only once. Then for each model, the \gls{gss}
considers only $\log_{1/\alpha}\left(\left|\mathcal{N}_{\text{x}}^{m}\right|-1\right)$
values for $n_{\text{x}}$, where where $\alpha$ is the golden ratio.
For each of these values, the optimal bit allocation for the quantization
of $\widehat{\boldsymbol{\theta}}$ has to be evaluated as well as
the residual $\boldsymbol{r}^{m}\left(\widehat{\boldsymbol{\theta}}_{\mathrm{q}}\left(n_{\text{x}}\right)\right)$.
Then, $L$ residual compression methods have to be compared. Consequently,
in a worst-case scenario, $\sum_{m=1}^{M}L\log_{1/\alpha}\left(\left|\mathcal{N}_{\text{x}}^{m}\right|-1\right)$
residual compression have to be performed. This is computationally
less complex than the \gls{es}. In practice, since $n_{\text{max}}$
is updated for each model, the size $\left|\mathcal{N}_{\text{x}}^{m}\right|$may
decrease with each tested model, thereby further reducing the overall
computational cost compared to the worst-case scenario.

\section{Using Rate-Distortion Models\label{sec:Distortion models for both stages}}

In this section, two models of the distortion $D^{m,\ell}\left(n_{\text{x}},n_{\text{r}}\right)$
introduced in \eqref{eq:Distorsion} are proposed. They are then exploited
to obtain an approximate, reduced-complexity, solution of \eqref{eq:optimization_problem}.

Section~\ref{subsec:Distortion Approximation with Gaussian Residuals}
introduces a first rate-distortion model assuming that the residuals
are uncorrelated and Gaussian. A second model is proposed in Section~\ref{subsec:Temporal Correlation in Residuals}
where temporal correlation of the residuals is taken into account.
Using these models, Section~\ref{subsec:Reduced-complexity search algorithm}
introduces an algorithm to select a subset of promising candidate
models for the first compression stage and to estimate optimal values
for $n_{\text{x}}$ and $n_{\text{r}}$ for each candidate model.

\subsection{Rate-Distortion Model Assuming Uncorrelated Gaussian Residuals\label{subsec:Distortion Approximation with Gaussian Residuals}}

Assume that the $N$ components of the residual $\boldsymbol{r}^{m}\left(\widehat{\boldsymbol{\theta}}_{\mathrm{q}}\left(n_{\text{x}}\right)\right)=\boldsymbol{x}-\boldsymbol{x}^{m}\left(\widehat{\boldsymbol{\theta}}_{\mathrm{q}}\left(n_{\text{x}}\right)\right)$
obtained at the output of the first stage are modeled as realizations
of independent and identically distributed zero-mean Gaussian variables
with variance $\left\Vert \boldsymbol{r}^{m}\left(\widehat{\boldsymbol{\theta}}_{\mathrm{q}}\left(n_{\text{x}}\right)\right)\right\Vert ^{2}/N$.
Moreover, assuming that the second stage achieves the rate-distortion
bound for Gaussian sources \cite{cover1999elements}, the distortion
of the output of the second stage when $n_{\text{r}}$ bits have been
allocated is 
\begin{equation}
D^{m,\ell}\left(n_{\text{x}},n_{\text{r}}\right)=\frac{1}{N}\left\Vert \boldsymbol{r}^{m}\left(\widehat{\boldsymbol{\theta}}_{\mathrm{q}}\left(n_{\text{x}}\right)\right)\right\Vert ^{2}2^{-2n_{\text{r}}/N}.\label{eq:Distorsion_Residual}
\end{equation}
The model \eqref{eq:Distorsion_Residual} and the following expressions
of the rate-distortion models do not account for the type of encoder
$\ell$ used at the second stage. Consequently, the superscript $\ell$
is omitted in $D^{m,\ell}\left(n_{\text{x}},n_{\text{r}}\right)$
in what follows.

The norm of $\boldsymbol{r}^{m}\left(\widehat{\boldsymbol{\theta}}_{\mathrm{q}}\left(n_{\text{x}}\right)\right)$
has to be evaluated for each $n_{\text{x}}$, which is one of the
sources of complexity of the algorithms presented in Section~\ref{sec:Proposed solutions}.
To address this issue, we first rewrite the residual as 
\begin{align}
\boldsymbol{r}^{m}\left(\widehat{\boldsymbol{\theta}}_{\mathrm{q}}\left(n_{\text{x}}\right)\right) & =\boldsymbol{x}-\boldsymbol{x}^{m}\left(\widehat{\boldsymbol{\theta}}\right)+\boldsymbol{x}^{m}\left(\widehat{\boldsymbol{\theta}}\right)-\boldsymbol{x}^{m}\left(\widehat{\boldsymbol{\theta}}_{\mathrm{q}}\left(n_{\text{x}}\right)\right)\\
 & =\boldsymbol{r}^{m}\left(\widehat{\boldsymbol{\theta}}\right)+\boldsymbol{e}_{\text{q}}^{m}(\widehat{\boldsymbol{\theta}},n_{\text{x}})\label{eq:ResTwoParts}
\end{align}
where $\boldsymbol{r}^{m}\left(\widehat{\boldsymbol{\theta}}\right)=\boldsymbol{x}-\boldsymbol{x}^{m}\left(\widehat{\boldsymbol{\theta}}\right)$
represents the error due to the inaccuracy of the model (without parameter
quantization) and $\boldsymbol{e}_{\text{q}}^{m}(\widehat{\boldsymbol{\theta}},n_{\text{x}})=\boldsymbol{x}^{m}(\widehat{\boldsymbol{\theta}})-\boldsymbol{x}^{m}\left(\widehat{\boldsymbol{\theta}}_{\mathrm{q}}\left(n_{\text{x}}\right)\right)$
represents the contribution to the residual of the model parameter
quantization error.

In general, one has
\begin{equation}
2\left|\langle\boldsymbol{r}^{m}\left(\widehat{\boldsymbol{\theta}}\right),\boldsymbol{e}_{\text{q}}^{m}(\widehat{\boldsymbol{\theta}},n_{\text{x}})\rangle\right|\ll\left\Vert \boldsymbol{r}^{m}\left(\widehat{\boldsymbol{\theta}}\right)\right\Vert ^{2}+\left\Vert \boldsymbol{e}_{\text{q}}^{m}(\widehat{\boldsymbol{\theta}},n_{\text{x}})\right\Vert ^{2},\label{eq:hyp_decorelation}
\end{equation}
so that $\boldsymbol{r}^{m}\left(\widehat{\boldsymbol{\theta}}\right)$
and $\boldsymbol{e}_{\text{q}}^{m}(\widehat{\boldsymbol{\theta}},n_{\text{x}})$
can be assumed as uncorrelated. Consequently, \eqref{eq:Distorsion_Residual}
can be rewritten as 
\begin{equation}
D^{m}\left(n_{\text{x}},n_{\text{r}}\right)=\frac{1}{N}\left(\left\Vert \boldsymbol{r}^{m}\left(\widehat{\boldsymbol{\theta}}\right)\right\Vert ^{2}+\left\Vert \boldsymbol{e}_{\text{q}}^{m}(\widehat{\boldsymbol{\theta}},n_{\text{x}})\right\Vert ^{2}\right)2^{-2n_{\text{r}}/N}.\label{eq:model_gauss_1}
\end{equation}

\subsection{Rate-Distortion Model Assuming Correlated Gaussian Residuals\label{subsec:Temporal Correlation in Residuals}}

In practice, for a signal $\boldsymbol{x}$ and a model $m$, the
successive samples of$\boldsymbol{r}^{m}\left(n_{\text{x}}\right)$
may exhibit some correlation. The aim of DCT, DWT, or VAE at the second
compression stage is to efficiently decorrelate the residual prior
to quantization and entropy coding. In this section, the temporal
correlation between the samples of both $\boldsymbol{r}^{m}\left(\widehat{\boldsymbol{\theta}}\right)$
and $\boldsymbol{e}_{\text{q}}^{m}(\widehat{\boldsymbol{\theta}},n_{\text{x}})$,
introduced in \eqref{eq:ResTwoParts}, is taken into account to get
a more accurate rate-distortion model.

To account for the temporal correlation between samples, the components
of the vector $\boldsymbol{r}^{m}\left(\widehat{\boldsymbol{\theta}}\right)=\left(r_{1}^{m}\left(\widehat{\boldsymbol{\theta}}\right),\dots,r_{N}^{m}\left(\widehat{\boldsymbol{\theta}}\right)\right)$
are modeled as realizations of an \gls{ar} process 
\begin{equation}
r_{n}^{m}\left(\widehat{\boldsymbol{\theta}}\right)=\sum_{p=1}^{P_{\text{m}}}\alpha_{\text{m},p}^{m}r_{n-p}^{m}\left(\widehat{\boldsymbol{\theta}}\right)+\varepsilon_{\text{m},n}^{m},
\end{equation}
where $P_{\text{m}}$ and $\alpha_{\text{m},p}^{m}$ represent the
order and coefficients of the \gls{ar} process and $\varepsilon_{\text{m},n}^{m}$
represents realizations of zero-mean uncorrelated Gaussian variables
with variance $\left(\sigma_{\text{m}}^{m}\right)^{2}$. Similarly,
for the components of $\boldsymbol{e}_{\text{q}}^{m}(\widehat{\boldsymbol{\theta}},n_{\text{x}})=(e_{\text{q},1}^{m}(\widehat{\boldsymbol{\theta}},n_{\text{x}}),\dots,e_{\text{q},N}^{m}(\widehat{\boldsymbol{\theta}},n_{\text{x}}))^{T}$,
the following model is considered 
\begin{equation}
e_{\text{q},n}^{m}(\widehat{\boldsymbol{\theta}},n_{\text{x}})=\sum_{p=1}^{P_{\text{q}}}\alpha_{\text{q},p}^{m}e_{\text{q},n-p}^{m}(\widehat{\boldsymbol{\theta}},n_{\text{x}})+\varepsilon_{\text{q},n}^{m}(n_{\text{x}}),
\end{equation}
where $P_{\text{q}}$ and $\alpha_{\text{q},p}^{m}$ represent the
order and coefficients of the \gls{ar} process, and the $\varepsilon_{\text{q},n}^{m}(\widehat{\boldsymbol{\theta}},n_{\text{x}})$
are realizations of zero-mean uncorrelated Gaussian noise with variance
$\left(\sigma_{\text{q}}^{m}\right)^{2}\left(n_{\text{x}}\right)$.
The variances $\left(\sigma_{\text{m}}^{m}\right)^{2}$ and $\left(\sigma_{\text{q}}^{m}\right)^{2}\left(n_{\text{x}}\right)$
are determined using estimates of the autocorrelation functions $\gamma_{\text{m}}^{m}(p)$
and $\gamma_{\text{q}}^{m}(p,n_{\text{x}})$ of $r_{n}^{m}\left(\widehat{\boldsymbol{\theta}}\right)$
and $e_{\text{q},n}^{m}(\widehat{\boldsymbol{\theta}},n_{\text{x}})$
via Yule-Walker equations \cite{walker1931periodicity} to obtain
\begin{equation}
\begin{pmatrix}\alpha_{\text{m},1}^{m}\\
\vdots\\
\alpha_{\text{m},P_{\text{m}}}^{m}\\
\left(\sigma_{\text{m}}^{m}\right)^{2}
\end{pmatrix}=\begin{pmatrix}\gamma_{\text{m}}^{m}(1) & \dots & \gamma_{\text{m}}^{m}(P_{\text{m}}) & 1\\
\gamma_{\text{m}}^{m}(0) & \dots & \gamma_{\text{m}}^{m}(P_{\text{m}}-1) & 0\\
\vdots & \ddots & \vdots & \vdots\\
\gamma_{\text{m}}^{m}(P_{\text{m}}-1) & \dots & \gamma_{\text{m}}^{m}(0) & 0
\end{pmatrix}^{-1}\begin{pmatrix}\gamma_{\text{m}}^{m}(0)\\
\vdots\\
\gamma_{\text{m}}^{m}(P_{\text{m}})
\end{pmatrix},\label{eq:sigma_m}
\end{equation}
and
\begin{equation}
\begin{pmatrix}\alpha_{\text{q},1}^{m}\\
\vdots\\
\alpha_{\text{q},P_{\text{q}}}^{m}\\
\left(\sigma_{\text{q}}^{m}\right)^{2}\left(n_{\text{x}}\right)
\end{pmatrix}=\begin{pmatrix}\gamma_{\text{q}}^{m}(1,\widehat{\boldsymbol{\theta}},n_{\text{x}}) & \dots & \gamma_{\text{q}}^{m}(P_{\text{q}},\widehat{\boldsymbol{\theta}},n_{\text{x}}) & 1\\
\gamma_{\text{q}}^{m}(0,\widehat{\boldsymbol{\theta}},n_{\text{x}}) & \dots & \gamma_{\text{q}}^{m}(P_{\text{q}}-1,\widehat{\boldsymbol{\theta}},n_{\text{x}}) & 0\\
\vdots & \ddots & \vdots & \vdots\\
\gamma_{\text{q}}^{m}(P_{\text{q}}-1,\widehat{\boldsymbol{\theta}},n_{\text{x}}) & \dots & \gamma_{\text{q}}^{m}(0,\widehat{\boldsymbol{\theta}},n_{\text{x}}) & 0
\end{pmatrix}^{-1}\begin{pmatrix}\gamma_{\text{q}}^{m}(0,n_{\text{x}})\\
\vdots\\
\gamma_{\text{q}}^{m}(P_{\text{q}},n_{\text{x}})
\end{pmatrix}.\label{eq:sigma_q}
\end{equation}

Then, assuming that the second compression stage is able to perfectly
decorrelate the components of $r_{n}^{m}\left(\widehat{\boldsymbol{\theta}}\right)$
and $e_{\text{q},n}^{m}(\widehat{\boldsymbol{\theta}},n_{\text{x}})$
so that only $\varepsilon_{\text{m},n}$ and $\varepsilon_{\text{q},n}(n_{\text{x}})$
have to be compressed, and that the second stage achieves the rate-distortion
bound for uncorrelated Gaussian sources, the distortion can be modeled
as
\begin{equation}
D^{m}\left(n_{\text{x}},n_{\text{r}}\right)=\left(\left(\sigma_{\text{m}}^{m}\right)^{2}+\left(\sigma_{\text{q}}^{m}\right)^{2}\left(n_{\text{x}}\right)\right)2^{-2n_{\text{r}}/N}.\label{eq:distortion_model}
\end{equation}
For example, if $\boldsymbol{r}^{m}\left(\widehat{\boldsymbol{\theta}}\right)$
and $\boldsymbol{e}_{\text{q}}^{m}(\widehat{\boldsymbol{\theta}},n_{\text{x}})$
are represented by first-order \gls{ar} models, \eqref{eq:distortion_model}
boils down to 
\begin{align}
D^{m}\left(n_{\text{x}},n_{\text{r}}\right) & \simeq\left(\frac{\widehat{\gamma}_{\text{m}}^{m}(0)^{2}-\widehat{\gamma}_{\text{m}}^{m}(1)^{2}}{\widehat{\gamma}_{\text{m}}^{m}(0)}+\frac{\widehat{\gamma}_{\text{q}}^{m}(0,\widehat{\boldsymbol{\theta}},n_{\text{x}})^{2}-\widehat{\gamma}_{\text{q}}^{m}(1,\widehat{\boldsymbol{\theta}},n_{\text{x}})^{2}}{\widehat{\gamma}_{\text{q}}^{m}(0,\widehat{\boldsymbol{\theta}},n_{\text{x}})}\right)2^{-2n_{\text{r}}/N}.\label{eq:distortion_model_1}
\end{align}

For a given vector $\boldsymbol{x}$ and model $m$, $\widehat{\boldsymbol{\theta}}$
has to be evaluated once. This is also the case for the biased estimate
\begin{equation}
\widehat{\gamma}_{\text{m}}^{m}(p)=\frac{1}{N}\sum_{n=p+1}^{N}r_{n}^{m}\left(\widehat{\boldsymbol{\theta}}\right)r_{n-p}^{m}\left(\widehat{\boldsymbol{\theta}}\right)\label{eq:Gammam_hat}
\end{equation}
of the autocorrelation function $\gamma_{\text{m}}^{m}(p)$ of $r_{n}^{m}\left(\widehat{\boldsymbol{\theta}}\right)$,
from which $\left(\sigma_{\text{m}}^{m}\right)^{2}$ is deduced using
\eqref{eq:sigma_m}. Nevertheless, $\boldsymbol{e}_{\text{q}}^{m}(\widehat{\boldsymbol{\theta}},n_{\text{x}})$
and the estimate of its autocorrelation function $\gamma_{\text{q}}^{m}(p,\widehat{\boldsymbol{\theta}},n_{\text{x}})$
have still to be evaluated for each $n_{\mathrm{x}}$. Proposition~\ref{prop:model_eq2}
simplifies the evaluation of $\widehat{\gamma}_{\text{q}}^{m}(p,\widehat{\boldsymbol{\theta}},n_{\text{x}})$.
\begin{prop}
\label{prop:model_eq2} Consider a model with output $\boldsymbol{x}^{m}\left(\boldsymbol{\theta}\right)$
and assume that the components of the quantization error $\boldsymbol{\varepsilon}\left(\widehat{\boldsymbol{\theta}},n_{\text{x}}\right)=\widehat{\boldsymbol{\theta}}-\widehat{\boldsymbol{\theta}}_{\mathrm{q}}\left(n_{\text{x}}\right)$
of the estimated parameter vector $\widehat{\boldsymbol{\theta}}$
are zero-mean and uncorrelated then
\begin{align}
\widehat{\gamma}_{\text{\ensuremath{\mathrm{q}}}}^{m}\left(p,\widehat{\boldsymbol{\theta}},n_{\mathrm{x}}\right) & =\sum_{k=1}^{K}h_{k}^{m}\left(p,\widehat{\boldsymbol{\theta}}\right)\sigma_{k}^{2}\left(\widehat{\boldsymbol{\theta}},n_{\text{x}}\right),\label{eq:model_e_q_2}
\end{align}
where
\begin{equation}
h_{k}^{m}\left(p,\widehat{\boldsymbol{\theta}}\right)=\frac{1}{N}\sum_{n=p+1}^{N}\frac{\partial x_{n}^{m}\left(\widehat{\boldsymbol{\theta}}\right)}{\partial\widehat{\theta}_{k}}\frac{\partial x_{n-p}^{m}\left(\widehat{\boldsymbol{\theta}}\right)}{\partial\widehat{\theta}_{k}}\label{eq:pk_j}
\end{equation}
and 
\begin{equation}
\sigma_{k}^{2}\left(\widehat{\boldsymbol{\theta}},n_{\text{x}}\right)=\mathbb{E}\left(\varepsilon_{k}^{2}\left(\widehat{\boldsymbol{\theta}},n_{\text{x}}\right)\right)\label{eq:VarianceEpsk}
\end{equation}
is the variance of the $k$-th component of $\boldsymbol{\varepsilon}\left(\widehat{\boldsymbol{\theta}},n_{\text{x}}\right)$.
\end{prop}
\begin{proof}
See Appendix~\ref{subsec:Proof of proposition model eq2}.
\end{proof}
The term $h_{k}(p,\widehat{\boldsymbol{\theta}})$ in \eqref{eq:pk_j}
is the expected value of the autocorrelation function of the sensitivity
of the model output with respect to the $k$-th component of $\widehat{\boldsymbol{\theta}}$.
The estimate $\widehat{\gamma}_{\text{q}}^{m}(p,\widehat{\boldsymbol{\theta}},n_{\text{x}})$
is then obtained as a weighted average of $h_{k}^{m}(p,\widehat{\boldsymbol{\theta}})$,
where the weights are given by the variances of the quantization error
of each component of the estimated parameter vector.

The following corollaries provide expressions of $h_{k}(p,\widehat{\boldsymbol{\theta}})$
for different models.
\begin{cor}
For a sinusoidal model with parameter vector $\widehat{\boldsymbol{\theta}}=\left(\widehat{a},\widehat{f},\widehat{\phi}\right)^{T}$,
\begin{align}
h_{1}(p,\widehat{\boldsymbol{\theta}}) & =\frac{1}{N}\sum_{n=p+1}^{N}\cos\left(2\pi\widehat{f}nT_{\text{s}}+\widehat{\phi}\right)\cos\left(2\pi\widehat{f}\left(n-p\right)T_{\text{s}}+\widehat{\phi}\right)\nonumber\\
 & =\frac{1}{2N}\left(\cos\left(2\pi\widehat{f}\left(N+1\right)T_{\text{s}}+2\widehat{\phi}\right)\frac{\sin\left(2\pi\widehat{f}\left(N-p\right)T_{\text{s}}\right)}{\sin\left(2\pi\widehat{f}T_{\text{s}}\right)}+\left(N-p\right)\cos\left(2\pi\widehat{f}pT_{\text{s}}\right)\right),\\
h_{2}(p,\widehat{\boldsymbol{\theta}}) & =\frac{4\pi^{2}T_{\text{s}}^{2}\widehat{a}^{2}}{N}\sum_{n=p+1}^{N}n\left(n-p\right)\sin\left(2\pi\widehat{f}nT_{\text{s}}+\widehat{\phi}\right)\sin\left(2\pi\widehat{f}\left(n-p\right)T_{\text{s}}+\widehat{\phi}\right),\\
h_{3}(p,\widehat{\boldsymbol{\theta}}) & =\frac{\widehat{a}^{2}}{N}\sum_{n=p+1}^{N}\sin\left(2\pi\widehat{f}nT_{\text{s}}+\widehat{\phi}\right)\sin\left(2\pi\widehat{f}\left(n-p\right)T_{\text{s}}+\widehat{\phi}\right)\nonumber\\
 & =\frac{\widehat{a}^{2}}{2N}\left(-\cos\left(2\pi\widehat{f}\left(N+1\right)T_{\text{s}}+2\widehat{\phi}\right)\frac{\sin\left(2\pi\widehat{f}\left(N-p\right)T_{\text{s}}\right)}{\sin\left(2\pi\widehat{f}T_{\text{s}}\right)}+\left(N-p\right)\cos\left(2\pi\widehat{f}pT_{\text{s}}\right)\right).
\end{align}
\end{cor}
\begin{proof}
See Appendix~\ref{subsec:gamma_sin}.
\end{proof}
\begin{cor}
For a Tchebychev polynomial model of degree $K-1$ and parameter vector
$\widehat{\boldsymbol{\theta}}$, 
\begin{equation}
	h_{k}(p,\widehat{\boldsymbol{\theta}})=\frac{1}{N}\sum_{n=p+1}^{N}\mathcal{T}_{k}\left(2\frac{n}{N}-1\right)\mathcal{T}_{k}\left(2\frac{n-p}{N}-1\right),
\end{equation}
for $k=0,\dots,K-1$.
\end{cor}
\begin{proof}
See Appendix~\ref{subsec:gamma_poly}.
\end{proof}
\begin{cor}
For a sample predictive model of order $K$ and offset $\eta$, 
\begin{equation}
h_{k}(p)=\frac{1}{N}\sum_{n=p+1}^{N}\widehat{x}_{(i-1)N+n-\eta-k+1}\widehat{x}_{(i-1)N+n-p-\eta-k+1},
\end{equation}
for $k=1,\dots,K$.
\end{cor}
\begin{proof}
See Appendix~\ref{subsec:gamma_samples_pred}.

For parameter predictive model described in Section~\ref{subsec:Optimal representation of the model parameters},
the entries of $\widehat{\boldsymbol{\delta}}$ are scalar quantized
to get $\widehat{\boldsymbol{\delta}}_{\text{q}}=\widehat{\boldsymbol{\delta}}+\boldsymbol{\varepsilon}$,
where $\boldsymbol{\varepsilon}$ is the quantization error vector.
Consequently, $h_{k}(p)$, $k=1,\dots,K$ can be evaluated considering
$\widehat{\boldsymbol{\delta}}$ instead of $\widehat{\boldsymbol{\theta}}$,
whatever the model.
\end{proof}

\subsection{Reduced-Complexity Search\label{subsec:Reduced-complexity search algorithm}}

The following reduced-complexity search algorithm exploits, for each
model $m\in\mathcal{M}$, the analytical model $D^{m}(n_{\text{x}},n_{\text{r}})$,
either given by \eqref{eq:model_gauss_1} or by \eqref{eq:distortion_model}
to get an approximate solution of \eqref{eq:optimization_problem}
by solving 
\begin{align}
\underline{n}_{\text{x}}^{m},\underline{n}_{\text{r}}^{m} & =\arg\min_{n_{\text{x}},n_{\text{r}}}n_{\text{h}}+n_{\text{x}}+n_{\text{r}}\label{eq:optimization_problem-1}\\
 & \text{s.t. }D^{m}\left(n_{\text{x}},n_{\text{r}}\right)\leqslant D_{\text{max}}.\nonumber 
\end{align}

Since $D^{m}(n_{\text{x}},n_{\text{r}})$ is only an approximation
of $D^{m,\ell}(n_{\text{x}},n_{\text{r}})$, an exhaustive or a \gls{gss} is then performed considering a subset $\mathcal{M}'\subset\mathcal{M}$
of most promising models, \emph{i.e.}, of models with the smallest
$\underline{n}_{\text{x}}^{m}+\underline{n}_{\text{r}}^{m}$. For
each of these models, a subset $\mathcal{N}_{\text{x}}^{'m}$ of values
around $\underline{n}_{\text{x}}^{m}$ is also considered to account
for the model inaccuracy.

Algorithm~\ref{alg:distortion_model} takes $\boldsymbol{x}$ as
input and outputs the subset $\mathcal{M}'$ and the set $\mathcal{N}_{\text{x}}^{'m}$
for each $m\in\mathcal{M}'$. For each model $\ensuremath{m}$ (line
1), $\widehat{\boldsymbol{\theta}}$ and $\boldsymbol{r}^{m}\left(\widehat{\boldsymbol{\theta}}\right)$
are evaluated at line 2 and the variance $\left(\sigma_{\text{m}}^{m}\right)^{2}$
using \eqref{eq:sigma_m} (line 3 and 4). The variable $n_{\text{max}}$,
which stores the minimum number of bits required for compression,
is initialized to infinity (line 5). For each value of $n_{\text{x}}$,
starting from zero (lines 6 and 7), then $\left(\sigma_{\text{q}}^{m}\right)^{2}\left(n_{\text{x}}\right)$
is determined using \eqref{eq:sigma_q} at line 8, as detailed in
Section~\ref{sec:Distortion models for both stages}. Based on this
value, the corresponding $n_{\text{r}}$ is obtained by solving \ref{eq:optimization_problem-1}
and imposing $D^{m}\left(n_{\text{x}},n_{\text{r}}\right)\leqslant D_{\text{max}}$
in \eqref{eq:distortion_model}, to get
\begin{equation}
n_{\text{r}}\left(n_{\text{x}}\right)=\left\lceil \frac{N}{2}\log_{2}\left(\frac{\left(\sigma_{\text{m}}^{m}\right)^{2}+\left(\sigma_{\text{q}}^{m}\right)^{2}\left(n_{\text{x}}\right)}{D_{\text{max}}}\right)\right\rceil ,
\end{equation}
see line 8. Assuming that $n_{\text{x}}+n_{\text{r}}(n_{\text{x}})$
is convex, if $n_{\text{x}}+n_{\text{r}}(n_{\text{x}})<n_{\text{max}}$,
$n_{\text{max}}$ is updated (line 10). The process then increments
$n_{\text{x}}$ (line 12), and the candidate output values $\underline{n}_{\text{x}}^{m}$
and $\underline{n}_{\text{r}}^{m}$ are updated with the values $n_{\text{x}}$
and $n_{\text{r}}$ (line 11). The process stops when $n_{\text{x}}+n_{\text{r}}(n_{\text{x}})\geqslant n_{\text{max}}$
(line 13), and the set 
\begin{equation}
\mathcal{N}_{\text{x}}^{'m}=\left[\underline{n}_{\text{x}}^{m}-\lfloor\Delta_{\text{nx}}/2\rfloor,\underline{n}_{\text{x}}^{m}+\lfloor\Delta_{\text{nx}}/2\rfloor\right],
\end{equation}
of candidates for $\widehat{n}_{\text{x}}$ is formed (line 14) to
perform some refined search around $\underline{n}_{\text{x}}^{m}$.

Once all models have been processed, a subset $\mathcal{M}'\subset\mathcal{M}$
containing the $\Delta_{\text{M}}\leqslant M$ most promising models
indexes is formed (line 18). A reduced complexity \gls{es}
or \gls{gss} may then be performed considering all $m\in\text{\ensuremath{\mathcal{M}}'}$
and all $n_{\text{x}}\in\mathcal{N}_{\text{x}}^{'m}$.

\begin{algorithm}[!h]
\caption{Preselection of candidate models and $n_{\text{x}}$ values}
\label{alg:distortion_model}

\For{each $m\in\mathcal{M}$}{

Evaluate $\widehat{\boldsymbol{\theta}}$ and $\boldsymbol{r}^{m}\left(\widehat{\boldsymbol{\theta}}\right)=\boldsymbol{x}-\boldsymbol{x}^{m}\left(\widehat{\boldsymbol{\theta}}\right)$
using \eqref{eq:estimation}\textbackslash ;

Evaluate an estimate $\widehat{\gamma}_{\text{m}}^{m}(p)$ of $\gamma_{\text{m}}^{m}(p)$
from $\boldsymbol{r}^{m}\left(\widehat{\boldsymbol{\theta}}\right)$\textbackslash ;

Evaluate $\left(\sigma_{\text{m}}^{m}\right)^{2}$ from $\widehat{\gamma}_{\text{m}}^{m}(p)$
using \eqref{eq:sigma_m}\textbackslash ;

$n_{\max}=\infty$\;

$n_{\text{x}}=0$\;

\While{true}{

Use offline estimate of $\left(\sigma_{\text{q}}^{m}\right)^{2}\left(n_{\text{x}}\right)$
using \eqref{eq:sigma_q} to get
\[
n_{\text{r}}\left(n_{\text{x}}\right)=\left\lceil \frac{N}{2}\log_{2}\left(\frac{\left(\sigma_{\text{m}}^{m}\right)^{2}+\left(\sigma_{\text{q}}^{m}\right)^{2}\left(n_{\text{x}}\right)}{D_{\text{max}}}\right)\right\rceil 
\]
tcc*{Solve \eqref{eq:optimization_problem-1} using \eqref{eq:distortion_model}}

\eIf{$n_{\text{x}}+n_{\text{r}}\left(n_{\text{x}}\right)<n_{\text{max}}$}{

$n_{\text{max}}=n_{\text{x}}+n_{\text{r}}\left(n_{\text{x}}\right)$\;

$\underline{n}_{\text{x}}^{m}=n_{\text{x}}$ ; $\underline{n}_{\text{r}}^{m}=n_{\text{r}}\left(n_{\text{x}}\right)$
\;

$n_{\text{x}}=n_{\text{x}}+1$ ; }{$\mathcal{N}_{\text{x}}^{'m}=\left[\underline{n}_{\text{x}}^{m}-\lfloor\frac{\Delta_{\text{nx}}}{2}\rfloor,\underline{n}_{\text{x}}^{m}+\lfloor\frac{\Delta_{\text{nx}}}{2}\rfloor\right]$
Break\;} } }

Build $\mathcal{M}'\subset\mathcal{M}$ of the $\Delta_{\text{M}}$
model indexes with smallest $\underline{n}_{\text{x}}^{m}+\underline{n}_{\text{r}}^{m}$.
\end{algorithm}

This approach reduces the computational cost by limiting the search
space for $n_{\text{x}}$. In a worst-case scenario, the number of
residual compression methods to evaluate is 
\begin{equation}
\sum_{m=1}^{M'}L|\mathcal{N}_{\text{x}}^{'m}|=\Delta_{\text{M}}L\Delta_{\text{nx}}
\end{equation}
when using \gls{es} (see Section~\ref{subsec:Exhaustive Search}),
and $\Delta_{\text{M}}L\log_{1/\alpha}(\Delta_{\text{nx}})$ with
the \gls{gss} (see Section~\ref{sec:Golden Section Search}).
The computational cost is controlled via the choice of $\Delta_{\text{M}}$
and $\Delta_{\text{nx}}$.

\section{Experiments\label{sec:Simulations examples}}

Section~\ref{subsec:Simulation set-up} introduces the simulation
setup. In Section~\ref{subsec:Hypothesis testing}, the validity
of several hypotheses considered in Sections~\ref{sec:Golden Section Search}, and \ref{sec:Distortion models for both stages}
are evaluated. Then, Section~\ref{subsec:Evaluation of Bit Allocation and Model Selection Using Reduced Complexity Search}
shows the performance of the proposed \gls{dm} in estimating $\ensuremath{\widehat{n}_{\text{x}}}$
and $\ensuremath{\widehat{n}_{\text{r}}}$ for a target distortion
$D_{\text{max}}$ for different values of $\Delta_{\text{nx}}$ and
$\Delta_{\text{M}}$. In Section~\ref{subsec:Comparison with Literature},
the compression performance for different values of $D_{\text{max}}$
and window sizes is compared with three reference methods considering
a large database of signal samples. Specifically, we analyze the \gls{es}
approach in Section~\ref{subsec:Exhaustive Search}, the \gls{gss}
approach in Section~\ref{sec:Golden Section Search}, and the \gls{es}
approach with \gls{dm} as well as the \gls{es} approach with \gls{gss}
in Section~\ref{subsec:Reduced-complexity search algorithm} \footnote{Codes are available at: \url{https://github.com/CorentinPresvots/MMC-with-Quality-Constraint}}.

\subsection{Simulation set-up\label{subsec:Simulation set-up}}

Twelve three-phase voltage signals from the digital-fault-recording-database
\cite{Database_RTE} acquired on the French electrical grid are first
considered. The selected signals correspond to indices 85, 91, 98,
176, 189, 195, 282, 287, 316, 337, 371, and 380, as they represent
typical faults. All signals last one second and are sampled at 6400
Hz (128 samples per nominal period $f_{\text{n}}=50$ Hz). The length
of the window is set to $N=128$ samples. Consequently, each signal
is partitioned into 50 non-overlapping windows of 20 ms duration,
leading to a total of $1800$ vectors to compress. More extensive
analyses have been performed on the signals with index 91 and $337$
from \cite{Database_RTE}.

Table~\ref{tab:index_m} summarizes the different models used in
the first and second stages of the considered \gls{mmc} approach.
For the polynomial model, we choose uniform prior distributions $p_{\boldsymbol{\theta}}$
over intervals with widths $w_{K^{m}}$ for the $k$-th coefficient
of the polynomial model $m$. The values of $w_{K^{m}}$ were determined
to gather 90~\% of the distribution of $\theta_{K^{m}}$, based on
an analysis of 20,000 signals of 128 samples from \cite{Database_RTE}.

\begin{table}[!h]
\centering{}%
\begin{tabular}{|c|c|c|}
\hline 
Model & Definitions & $p_{\boldsymbol{\theta}}$\tabularnewline
\hline 
\hline 
Bypass &  & \tabularnewline
\hline 
Sinusoidal & \ref{eq:sin} & $\mathcal{U}\left(\boldsymbol{\theta};\begin{pmatrix}0.5\\
40.9\\
-\pi
\end{pmatrix},\begin{pmatrix}1\\
50.1\\
\pi
\end{pmatrix}\right)$\tabularnewline
\hline 
Poly. of order $K\in\left[0,\dots,9\right]$ & \ref{eq:poly} & $\mathcal{U}\left(\boldsymbol{\theta};-\begin{pmatrix}\frac{w_{1}^{m}}{2}\\
\dots\\
\frac{w_{K+1}^{m}}{2}
\end{pmatrix},\begin{pmatrix}\frac{w_{1}^{m}}{2}\\
\dots\\
\frac{w_{K+1}^{m}}{2}
\end{pmatrix}\right)$ \tabularnewline
\hline 
Sample pred., order $K\in\left[1,2\right]$, $\eta=0$ & \ref{eq:samples_pred} & $\mathcal{U}\left(\boldsymbol{\theta};-\frac{1}{2}\boldsymbol{1},\frac{1}{2}\boldsymbol{1}\right)$\tabularnewline
\hline 
Parameter pred. & \ref{eq:para_pred} & $\mathcal{U}\left(\boldsymbol{\delta};-0.1\cdot\boldsymbol{1},0.1\cdot\boldsymbol{1}\right)$ \tabularnewline
\hline 
\end{tabular}\caption{\gls{mmc} approach: Model types and \textit{a priori} distributions
for the model parameters considered in the first stage; $\mathcal{U}\left(\boldsymbol{\theta};\boldsymbol{a},\boldsymbol{b}\right)$
represents a uniform distribution over the box with center $0.5\left(\boldsymbol{a}+\boldsymbol{b}\right)$
with widths $\boldsymbol{b}-\boldsymbol{a}$; The vector $\boldsymbol{1}$
is a vector of ones of the same dimension as $\boldsymbol{\theta}$.}
\label{tab:index_m}
\end{table}

In the second compression stage, two scalable methods for residual
compression are considered, namely \gls{dct} followed by the \gls{bpc}
described in \cite{hoang_new_2009} and \gls{dwt} followed by the
\gls{ezw} approach of \cite{khan_embedded-zerotree-wavelet-based_2015}.

The preamble of each compressed data packet is organized as follows.
To represent one of the 15 model indexes, $n_{\text{m}}=4$ bits are
required, and $n_{\ell}=1$ bit is used to indicate the residual compression
method. The number of bits $n_{\text{nx}}$ indicating the length
of the model parameter field depends on the chosen model $m$ and
its number of parameters $K^{m}$. We limit the number of bits per
parameter to 12; consequently, $n_{\text{nx}}=\left\lceil \log_{2}\left(12K^{m}\right)\right\rceil $,
where $\left\lceil \cdot\right\rceil $ represents rounding upwards.
If the first compression stage is bypassed, no bits are required to
represent $n_{\text{x}}$.

The number of bits $n_{\text{nr}}$ indicating the length of the residual
field depends on the chosen model $\ell$. If the second stage is
bypassed, no bits are required to represent $n_{\text{r}}$ and $n_{\text{kr}}$.
The number of bits for each parameter is determined from the model
index $m$ and $n_{\text{x}}$, as detailed in Section~\ref{subsec:Optimal representation of the model parameters}.

For the number of bits $n_{\text{kx}}$ required to represent the
first scaling factor $k_{\text{x}}$, we assume that the absolute
value of the phase-to-ground voltage is between $0.5$~V and $500$~kV.
Since $2^{-k_{\text{x}}}x\in\left[0.5,1\right[$ for all $x\in\left[0.5,5\cdot10^{5}\right]$,
it follows that $k_{\text{x}}\in\mathbb{K}_{\text{x}}=\left[0,\dots,19\right]$,
leading to $n_{\text{kx}}=5$ bits.

If the maximum absolute value of the residual is less than $10^{-3}$,
the residual is not encoded. Since the maximum amplitude of the scaled
signal is in $\left[0.5,1\right[$, a relative error of less than
$0.1\%$ of nominal voltage is obtained,
which corresponds to the usual precision class of transformer instruments
described in the \gls{iec} 60044-1 standard \cite{IEC_60044_1},
\emph{i.e.}, less than $\pm90$ V.

For the number of bits $n_{\text{kr}}$ required to represent the
second scaling factor $k_{\text{r}}$, we require that $2^{-k_{\text{r}}}r\in\left[0.5,1\right[$
for all $r\in\left[10^{-3},1\right]$. This leads to $k_{\text{r}}\in\mathbb{K}_{\text{r}}=\left[-9,\dots,0\right]$
and $n_{\text{kr}}=4$ bits.

\subsection{Analysis of the considered hypotheses\label{subsec:Hypothesis testing}}

\begin{figure}[H]
\centering
\includegraphics[width=0.3\columnwidth]{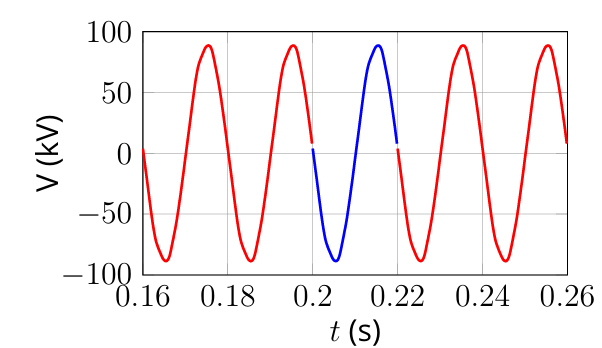} \includegraphics[width=0.3\columnwidth]{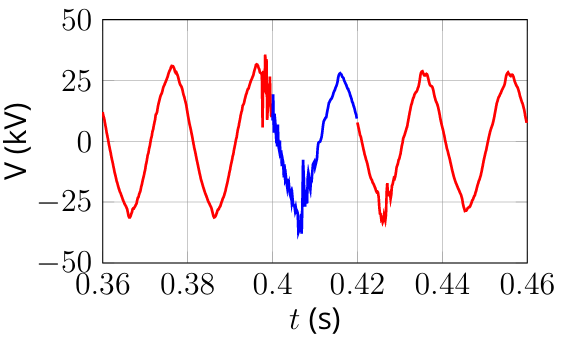}
\includegraphics[width=0.3\columnwidth]{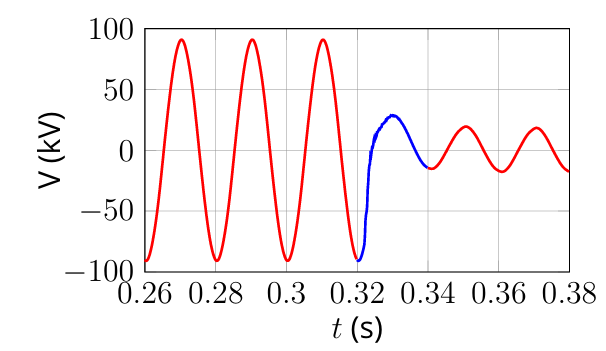} 

\caption{Three test signals of 128 samples (blue curves, sampling rate $f_{\text{s}}=6400$
Hz) extracted from the database \cite{Database_RTE}; The first two
signals are from the first phase of the signal with index 91, and
the third from the second phase of the signal with index 337.}
\label{fig_ch7:test_signal}
\end{figure}

Figure~\ref{fig_ch7:test_signal} illustrates three test signals
in blue, each containing 128 samples corresponding to 20 ms of data
extracted from the signals with index 91 and 337 in \cite{Database_RTE}.
These signals are representative of transient behaviors. They are
considered to evaluate the validity of various hypothses considered
in Sections~\ref{sec:Distortion models for both stages}, \ref{sec:Golden Section Search},
and \ref{sec:Distortion models for both stages}.

In the following section, we consider $D_{\text{max}}=200^{2}\,\text{V}^{2}$.

\subsubsection{Convexity of $n_{\text{x}}+n_{\text{r}}(n_{\text{x}})$ as a function
of $n_{\text{x}}$ under a constraint $D_{\text{max}}$}

\begin{figure}[H]
\centering
\includegraphics[width=0.33\columnwidth]{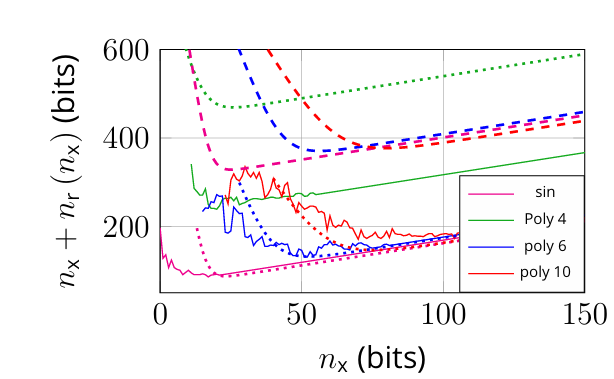}\includegraphics[width=0.33\columnwidth]{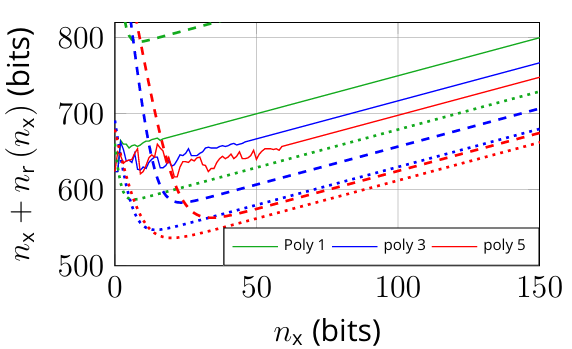}\includegraphics[width=0.33\columnwidth]{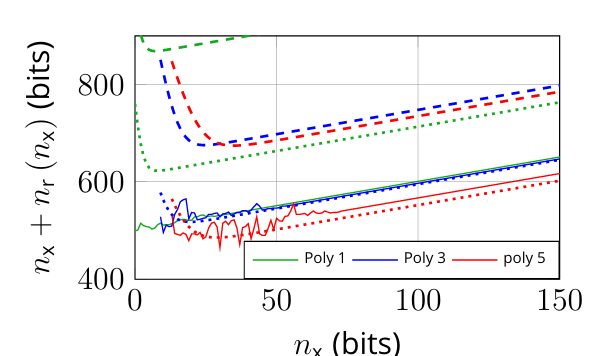}

\caption{Evolution of $n_{\text{x}}+n_{\text{r}}(n_{\text{x}})$ obtained by
\gls{es} (plain), the model \eqref{eq:model_gauss_1} (dashed),
and \eqref{eq:distortion_model_1} (dotted), as a function of $n_{\text{x}}$
using sinusoidal (pink), polynomial models of degree 4 (green), 6
(blue), and 10 (red) with test signal 1 (left), and polynomial models
of degree 1 (green), 3 (blue), and 5 (red) with test signals 2 (middle)
and 3 (right) under a distortion constraint of $D_{\text{max}}=200^{2}$
$\text{V}^{2}$.}
\label{fig:n_tot_n_x}
\end{figure}

In Section~\ref{sec:Golden Section Search}, we assume that $n_{\text{x}}+n_{\text{r}}(n_{\text{x}})$
is convex with respect to $n_{\text{x}}$ to apply the \gls{gss}
and find $\widehat{n}_{\text{x}}$. Figure~\ref{fig:n_tot_n_x} represents
$n_{\text{x}}+n_{\text{r}}(n_{\text{x}})$ as a function of $n_{\text{x}}$
obtained by an \gls{es} (plain) and by the models \eqref{eq:model_gauss_1}
(dashed) and \eqref{eq:distortion_model_1} (dotted) of $n_{\text{x}}+n_{\text{r}}(n_{\text{x}})$,
considering the three test signals and different signal models. Only
the global shape is convex, but there are irregularities, especially
at low bit rates. This is mainly due to the fact that adding one extra
bit to represent the model parameters does not always yield a residual
easier to encode at the second stage. These irregularities lead to
degraded performance of the \gls{gss} compared to the \gls{es},
as will be seen in Section~\ref{subsec:Comparison with Literature}.

The model \eqref{eq:distortion_model_1} provides usually better results
than \eqref{eq:model_gauss_1}. Even if there is some bias in the
value of $n_{\text{x}}+n_{\text{r}}(n_{\text{x}})$, the minimum obtained
considering \eqref{eq:distortion_model_1} is closer to the minimum
obtained by \gls{es} than that obtained using \eqref{eq:model_gauss_1}. 

\subsubsection{Evaluation of \eqref{eq:model_gauss_1}}

\begin{figure}[!h]
\centering
\includegraphics[width=0.8\columnwidth]{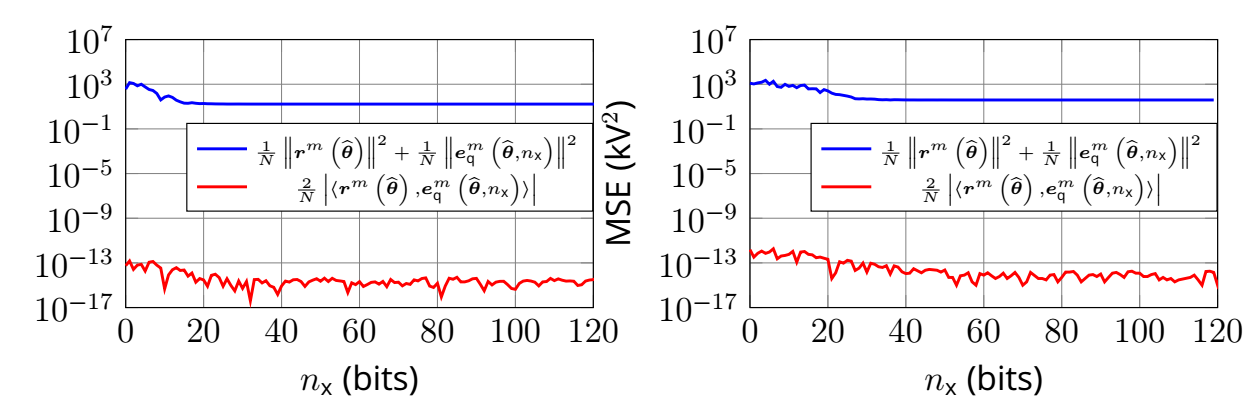}

\caption{Evolution of $\frac{1}{N}\left\Vert \boldsymbol{r}^{m}\left(\widehat{\boldsymbol{\theta}}\right)\right\Vert ^{2}+\frac{1}{N}\left\Vert \boldsymbol{e}_{\text{q}}^{m}(\widehat{\boldsymbol{\theta}},n_{\text{x}})\right\Vert ^{2}$
(blue) and $\frac{2}{N}\left|\langle\boldsymbol{r}^{m}\left(\widehat{\boldsymbol{\theta}}\right),\boldsymbol{e}_{\text{q}}^{m}(\widehat{\boldsymbol{\theta}},n_{\text{x}})\rangle\right|$
(red), both as a function of $n_{\text{x}}$, using a 3th-degree polynomial
model for the second test signal (middle) and a 5rd-degree polynomial
for the third test signal (right) shown in Figure~\ref{fig_ch7:test_signal}.}
\label{fig:test_signal_uncorrelated}
\end{figure}

In Section~\ref{subsec:Distortion Approximation with Gaussian Residuals},
we assume that 
\begin{equation}
2\left|\langle\boldsymbol{r}^{m}\left(\widehat{\boldsymbol{\theta}}\right),\boldsymbol{e}_{\text{q}}^{m}(\widehat{\boldsymbol{\theta}},n_{\text{x}})\rangle\right|\ll\left\Vert \boldsymbol{r}^{m}\left(\widehat{\boldsymbol{\theta}}\right)\right\Vert ^{2}+\left\Vert \boldsymbol{e}_{\text{q}}^{m}(\widehat{\boldsymbol{\theta}},n_{\text{x}})\right\Vert ^{2},
\end{equation}
which underlies \eqref{eq:model_gauss_1}. For the two considered
test signals, Figure~\ref{fig:test_signal_uncorrelated} shows that
this condition holds for all values of $n_{\text{x}}$.

\subsubsection{Evaluation of \eqref{eq:distortion_model}}

\begin{figure}[!h]
\centering
\includegraphics[width=0.8\columnwidth]{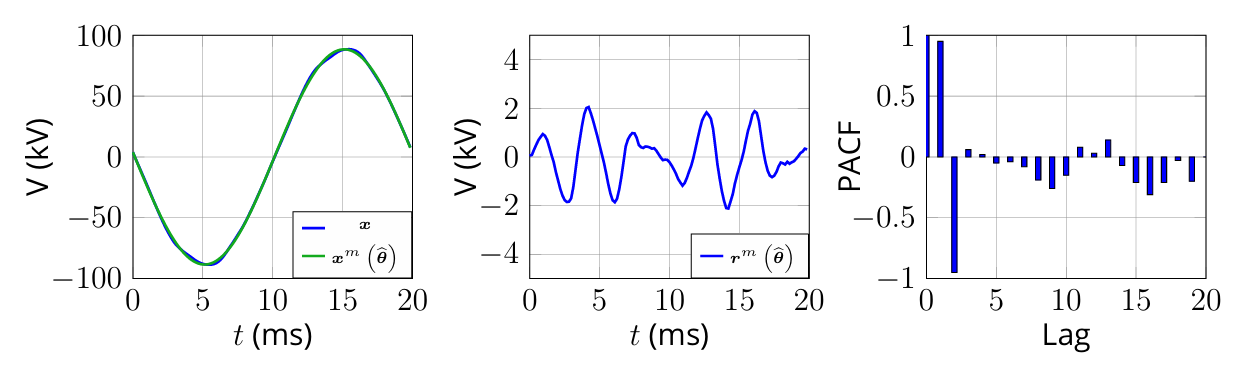}

\caption{For test signal 1 shown in Figure~\ref{fig_ch7:test_signal}. Left:
The test signal (blue) and its reconstruction using a sinusoidal model
with unquantized parameters (green). Middle: $\boldsymbol{r}^{m}\left(\widehat{\boldsymbol{\theta}}\right)$.
Right: \gls{pacf} of $\boldsymbol{r}^{m}\left(\widehat{\boldsymbol{\theta}}\right)$.}
\label{fig:test_signal_em-1}
\end{figure}

\begin{figure}[htpb]
\centering
\includegraphics[width=0.8\columnwidth]{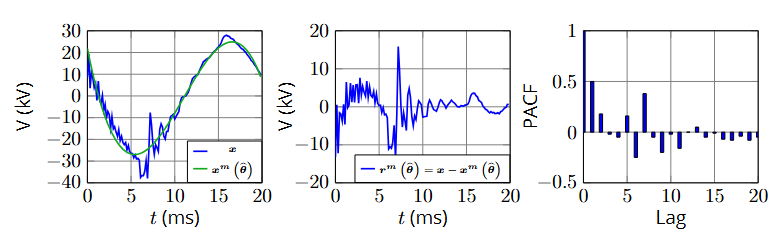}
\caption{For test signal 3 shown in Figure~\ref{fig_ch7:test_signal}: test
signal in blue and its reconstruction using a 3rd-degree polynomial
model with unquantized parameters in green (left), $\boldsymbol{r}^{m}\left(\widehat{\boldsymbol{\theta}}\right)$
(middle), and \gls{pacf} of $\boldsymbol{r}^{m}\left(\widehat{\boldsymbol{\theta}}\right)$
(right).}
\label{fig:test_signal_em_2} 
\end{figure}

In Section~\ref{subsec:Temporal Correlation in Residuals}, we assume
that the components of the estimation residual $\boldsymbol{r}^{m}\left(\widehat{\boldsymbol{\theta}}\right)$
may exhibit temporal correlation. Figure~\ref{fig:test_signal_em-1}
shows $\boldsymbol{r}^{m}\left(\widehat{\boldsymbol{\theta}}\right)$
and its \gls{pacf} for the first test signal and a sinusoidal model,
indicating a significant correlation. Figure~\ref{fig:test_signal_em_2}
and \ref{fig:test_signal_em_2} present the corresponding results
for the second and third test signals, using polynomial models of
degree 5 and 3, respectively.

\begin{figure}[htpb]
\centering
\includegraphics[width=0.8\columnwidth]{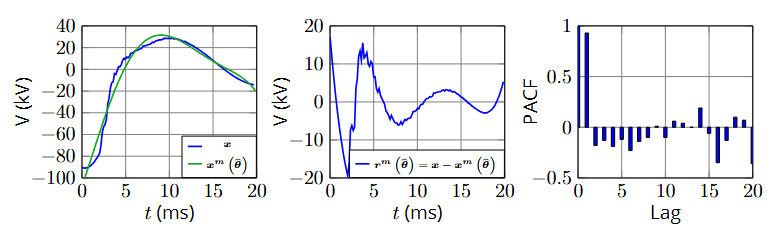}
\caption{For test signal 3 shown in Figure~\ref{fig_ch7:test_signal}: test
signal in blue and its reconstruction using a 5th-degree polynomial
model with unquantized parameters in green (left), $\boldsymbol{r}^{m}\left(\widehat{\boldsymbol{\theta}}\right)$
(middle), and \gls{pacf} of $\boldsymbol{r}^{m}\left(\widehat{\boldsymbol{\theta}}\right)$
(right).}
\label{fig:test_signal_em} 
\end{figure}

\begin{figure}[htpb]
\begin{centering}
\includegraphics[width=0.8\columnwidth]{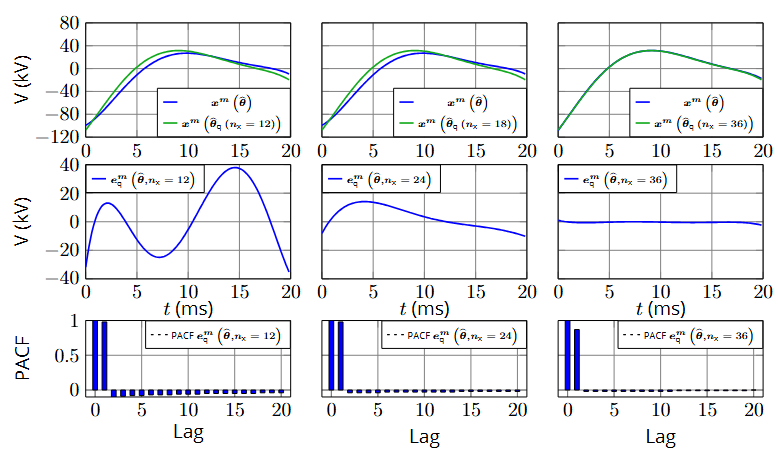}
\par\end{centering}
\caption[Example of temporal correlation in residual as a function of $n_{\text{x}}$]{First row: Reconstructed signal 1 (see Figure~\ref{fig_ch7:test_signal})
using a polynomial model of degree 5 without parameter quantization
(green), and with parameters quantized with $n_{\text{x}}=12$ bits
(left), $24$ bits (middle), and $36$ bits (right) (blue). Second
row: Quantization error $\boldsymbol{e}_{\text{q}}^{m}(\widehat{\boldsymbol{\theta}},n_{\text{x}})$
for parameter quantization with $n_{\text{x}}=12$ bits (left), $24$
bits (middle), and $36$ bits (right). Third row: \gls{pacf} of
$\boldsymbol{e}_{\text{q}}^{m}(\widehat{\boldsymbol{\theta}},n_{\text{x}})$
with parameters quantized with $n_{\text{x}}=12$ bits (left), $24$
bits (middle), and $36$ bits (right).}
\label{fig:test_signal_eq} 
\end{figure}

\begin{figure}[htp]
\begin{centering}
\includegraphics[width=0.8\columnwidth]{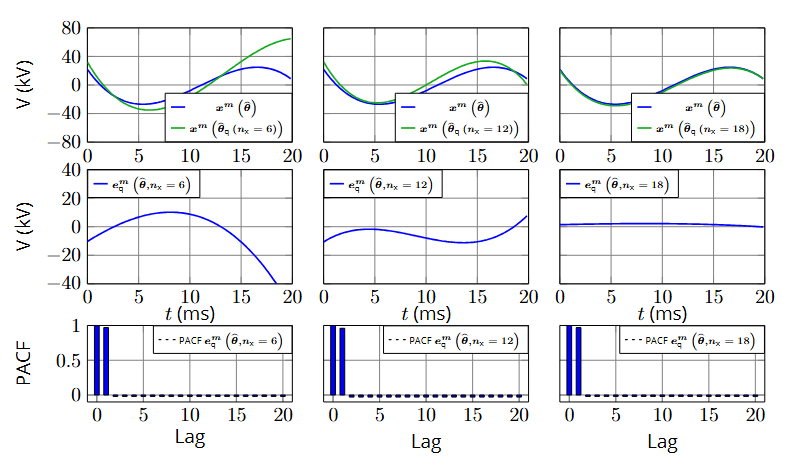}
\par\end{centering}
\caption{First row: Reconstructed signal 2 (see Figure~\ref{fig_ch7:test_signal})
using a polynomial of degree 3 without parameter quantization (green),
and with parameters quantized to 6 bits (left), 12 bits (middle),
and 18 bits (right) (blue). Second row: Quantization error $\boldsymbol{e}_{\text{q}}^{m}(\widehat{\boldsymbol{\theta}},n_{\text{x}})$
with parameter quantization at 6 bits (left), 12 bits (middle), and
18 bits (right). Third row: \gls{pacf} of $\boldsymbol{e}_{\text{q}}^{m}(\widehat{\boldsymbol{\theta}},n_{\text{x}})$
with parameters quantization at 6 bits (left), 12 bits (middle), and
18 bits (right).}
\label{fig:test_signal_eq_2} 
\end{figure}

Similarly, the components of the quantization error $\boldsymbol{e}_{\text{q}}^{m}(\widehat{\boldsymbol{\theta}},n_{\text{x}})$
are correlated, as seen in Figures~\ref{fig:test_signal_eq} and
\ref{fig:test_signal_eq_2}. This is confirmed by the \gls{pacf}
for different values of $n_{\text{x}}$ which is significant at lag
1.

\subsection{Efficiency of the bit allocation and model selection\label{subsec:Evaluation of Bit Allocation and Model Selection Using Reduced Complexity Search}}

\begin{figure}[htpb]
\begin{centering}
\includegraphics[width=0.8\columnwidth]{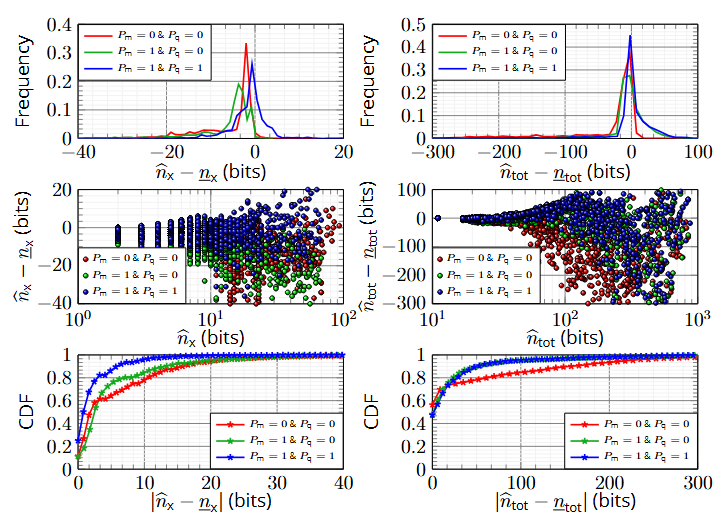}
\par\end{centering}
\caption{First row: Histograms of the $\widehat{n}_{\text{x}}-\underline{n}_{\text{x}}$(left)
and $\widehat{n}_{\text{tot}}-\underline{n}_{\text{tot}}$ (right),
where $\widehat{n}_{\text{x}}$ and $\widehat{n}_{\text{tot}}$ are
obtained with using \gls{es} with Algorithm~\ref{alg:optimal_solution},
and $\underline{n}_{\text{x}}$ and $\underline{n}_{\text{tot}}=\underline{n}_{\text{x}}+\underline{n}_{\text{r}}$
are obtained with \gls{dm} with Algorithm~\ref{alg:distortion_model},
for different values of $P_{\text{m}}$ and $P_{\text{q}}$ based
on the distortion model \eqref{eq:distortion_model}. Second row:
Scatter plot showing the distribution of $\widehat{n}_{\text{x}}-\underline{n}_{\text{x}}$
(left) and $\widehat{n}_{\text{tot}}-\underline{n}_{\text{tot}}$
(right). Third row: \gls{cdf} of the distribution of $\left|\widehat{n}_{\text{x}}-\underline{n}_{\text{x}}\right|$
(left) and $\left|\widehat{n}_{\text{tot}}-\underline{n}_{\text{tot}}\right|$
(right). Results are for $D_{\text{max}}=200^{2}\ V^{2}$.}

\label{fig:hist_nx}
\end{figure}

Figure~\ref{fig:hist_nx} (left) shows the error between $\widehat{n}_{\text{x}}$
obtained using \gls{es} with Algorithm~\ref{alg:optimal_solution}
and $\underline{n}_{\text{x}}$ predicted by \gls{dm} using Algorithm~\ref{alg:distortion_model}
when $D_{\text{max}}=200^{2}\ V^{2}$. Figure~\ref{fig:hist_nx}
(right) shows the error between $\widehat{n}_{\text{tot}}=\widehat{n}_{\text{x}}+\widehat{n}_{\text{r}}$
obtained using \gls{es} and $\underline{n}_{\text{tot}}=\underline{n}_{\text{x}}+\underline{n}_{\text{r}}$
predicted by \gls{dm}. Different orders $P_{\text{m}}$ and $P_{\text{q}}$
of the \gls{ar} model for the model error and quantization error
are considered.

In Figure~\ref{fig:hist_nx}, the histograms (top row) and the \glspl{cdf}
(bottom row) alongside show that a first-order \gls{ar} model of
the residuals of the first stage enhance the prediction accuracy of
$\widehat{n}_{\text{x}}$ using $\underline{n}_{\text{x}}$ (left).
For the prediction of $\widehat{n}_{\text{tot}}$ using $\underline{n}_{\text{tot}}$
(right), we observe that considering $P_{\text{m}}=1$ significantly
reduces the number of cases $\underline{n}_{\text{tot}}$ overestimates
$\widehat{n}_{\text{tot}}$. Using $P_{\text{q}}=1$ instead of $P_{\text{q}}=0$,
does not provide significant improvement.

The scatter plots of Figure~\ref{fig:hist_nx} (middle row) show
that larger prediction errors are generally associated with signals
requiring a high bit-rate to reach the target distortion. For the
right scatter plot, we observe that the models with $P_{\text{m}}=1$,
$P_{\text{q}}=1$ yields a less biased estimate of $\widehat{n}_{\text{x}}$.

\begin{figure}[H]
\centering{}\includegraphics[width=\columnwidth]{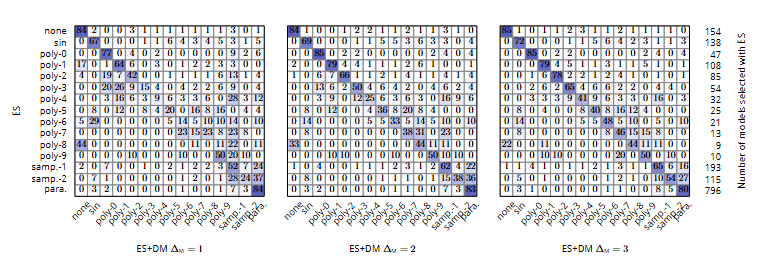}\caption[Evaluation of model selection estimation using rate-distortion models]{Confusion matrices showing the percentage of model selection of the \gls{es} with \gls{dm} compared to the \gls{es}
approach when $\Delta_{\text{M}}=1$ (left), $\Delta_{\text{M}}=2$
(middle), and $\Delta_{\text{M}}=3$ (right). The number of selection
of each model by the \gls{es} approach is on the right. Results
are based on 1800 vectors of 128 samples from \cite{Database_RTE}
with $D_{\text{max}}=200^{2}\ V^{2}$.}
\label{fig:confusion} 
\end{figure}

Initially, the set $\mathcal{M}$ of $14$ candidate models, see Table~\ref{tab:index_m}.
We compare the accuracy of model selection between the \gls{es}
approach (Section~\ref{subsec:Exhaustive Search}) and the \gls{es}
with \gls{dm} (Section~\ref{subsec:Reduced-complexity search algorithm}).
The impact of the number $\Delta_{\text{M}}$ of models considered
in the subset $\mathcal{M}'$ of promising models evaluated by Algorithm~\ref{alg:distortion_model},
used to initialize the set of models used by Algorithm~\ref{alg:optimal_solution}
For the \gls{es} approach, Fig.~\ref{fig:confusion} presents three confusion matrices for different values of $\Delta_{\text{M}}$. We can observe that increasing the number of retained models $\Delta_{\text{M}}$ increases the percentage of model selections by the \gls{es} with DM approach which are similar to that of the \gls{es} approach. 
The \gls{es} approach achieves $87.6$~bits per signal vector in average
for a distortion constraint of $D_{\text{max}}=200^{2}\,\text{V}^{2}$.
The model prediction performance improves when $\Delta_{\text{M}}$
increases. The average total bit rate also decreases from 92.6 bits
per signal for $\Delta_{\text{M}}=1$ to $90.4$~bits for $\Delta_{\text{M}}=2$,
and to $89.5$~bits for $\Delta_{\text{M}}=3$. By considering only 3
models out of the initial 14 for each signal, the reduced-complexity
\gls{dm} approach yields a computational cost reduction by a factor
of at least $4$ with an increase of the required rate of less than 2
bits per signal compared to \gls{es}.

Table~\ref{tab:rate_mrd_deltanx_delta_m} provides the average total
bit rate obtained by the \gls{es} with \gls{dm} to get a distortion
below $D_{\text{max}}=200$ $V^{2}$ for different values of $\Delta_{\text{M}}$
and $\Delta_{\text{nx}}$, illustrating the compromise between compression
efficiency and complexity. Considering $\Delta_{\text{M}}=3$ yields
a significant complexity reduction with an increase of less than $10\%$
of the required rate, whatever the value of $\Delta_{\text{nx}}$.

\begin{table}[!h]
\centering{}%
\begin{tabular}{|c|c|c|c|c|c|c|}
\hline 
\diagbox{$\Delta_{\text{M}}$}{$\Delta_{\text{nx}}$}  & 1  & 3  & 5  & 7  & 9  & 33\tabularnewline
\hline 
1  & 101.7  & 97.2  & 95.6  & 94.6  & 94.2  & 92.6\tabularnewline
\hline 
2  & 98.2  & 94.2  & 93.0  & 92.4  & 91.9  & 90.4 \tabularnewline
\hline 
3  & 96.7  & 93.4  & 91.9  & 91.4  & 90.9  & 89.5\tabularnewline
\hline 
14  & 94.3  & 91.7  & 90.1  & 89.5  & 89.16  & 87.9\tabularnewline
\hline 
\end{tabular}\caption{Average total bit rate (bits/window) achieved with the \gls{es}
with \gls{dm} approach considering \eqref{eq:distortion_model}
with $P_{\text{m}}=1$ and $P_{\text{q}}=1$ for different values
of $\Delta_{\text{M}}$ and $\Delta_{\text{nx}}$. Results are averaged
on 1800 vectors of 128 samples from digital-fault-recorded database
\cite{Database_RTE} with $D_{\text{max}}=200^{2}\ V^{2}$.}
\label{tab:rate_mrd_deltanx_delta_m} 
\end{table}

\subsection{Comparison with alternative approaches\label{subsec:Comparison with Literature}}

The proposed approaches are compared to reference methods. The first
has been proposed in \cite{sabarimalai_manikandan_simultaneous_2015}
and involves a \gls{dwt} followed by the \gls{ezw} entropy coder
\footnote{Code available at this link: \url{https://github.com/CorentinPresvots/MMC-with-Rate-Constraint}}.
The second is the technique from \cite{sabarimalai_manikandan_simultaneous_2015},
which is based on a sparse representation using \gls{ohd} \footnote{Code available at this link: \url{https://github.com/CorentinPresvots/OHD}}.
The third is the compression method from \cite{nascimento_improved_2020},
which employs \gls{wsqm} \footnote{Code available at this link: \url{https://github.com/CorentinPresvots/WSQM}}.
The \gls{wsqm} approach considering both linear and exponential decay
models for spectral shape, balancing efficiency and complexity.

In this section, we first evaluate the impact of $D_{\text{max}}$
and of the window size $N$ on the bitrate and on the compression
efficiency. Then their impact on the computational cost is evaluated.

\begin{figure}[htpb]
\centering
\includegraphics[width=0.9\columnwidth]{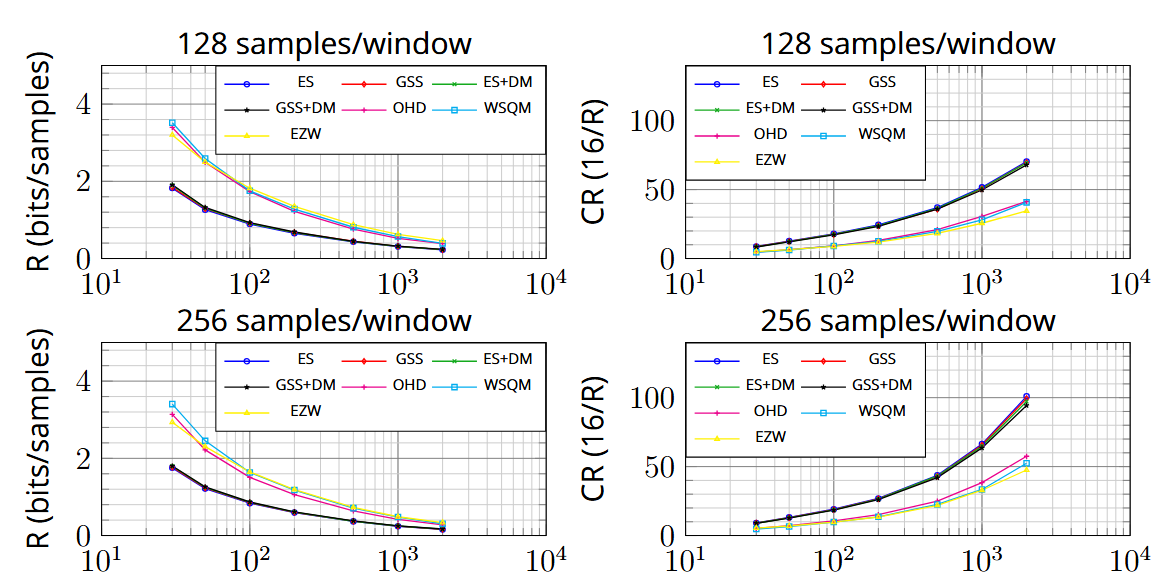}
\par
\includegraphics[width=0.9\columnwidth]{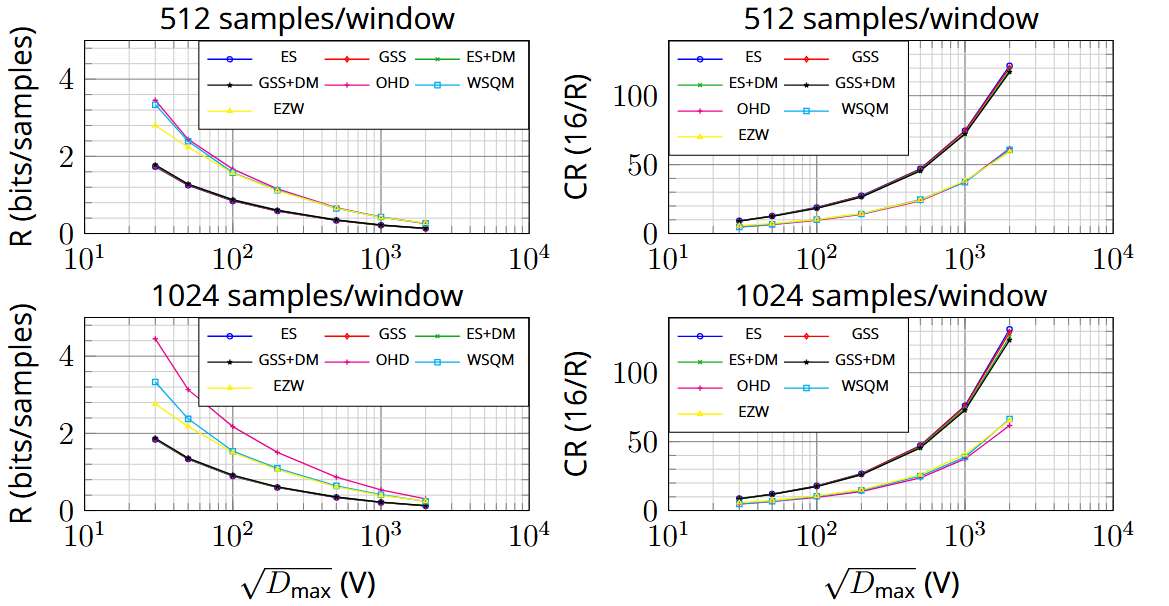}

\caption{Bitrate (left) and \gls{cr} (right) as a function of $\sqrt{\ensuremath{D_{\max}}}$
for \gls{es}, \gls{es} with \gls{dm}, \gls{gss}, \gls{gss} with
\gls{dm}, \gls{wsqm}, \gls{ohd}, and \gls{ezw}. Rows correspond
to window sizes of 128, 256, 512, and 1024 samples, respectively.
Results are averaged over the twelve first-phase voltage signals described
in Section~\ref{subsec:Simulation set-up}.}
\label{fig:Comp}
\end{figure}

Figure~\ref{fig:Comp} shows that the proposed approaches (\gls{es},
\gls{es} with \gls{dm}, \gls{gss}, and \gls{gss} with \gls{dm})
achieve the best performance in terms of bitrate for different distortion
constraints $D_{\text{max}}$, outperforming the reference methods
(\gls{wsqm}, \gls{ohd}, and \gls{ezw}). The proposed methods perform
quite similarly, even if \gls{es} provides the best results, followed
by \gls{gss}, then \gls{es} with \gls{dm}, and finally \gls{gss}
with \gls{dm}.

\begin{figure}[htpb]
\centering
\includegraphics[width=0.9\columnwidth]{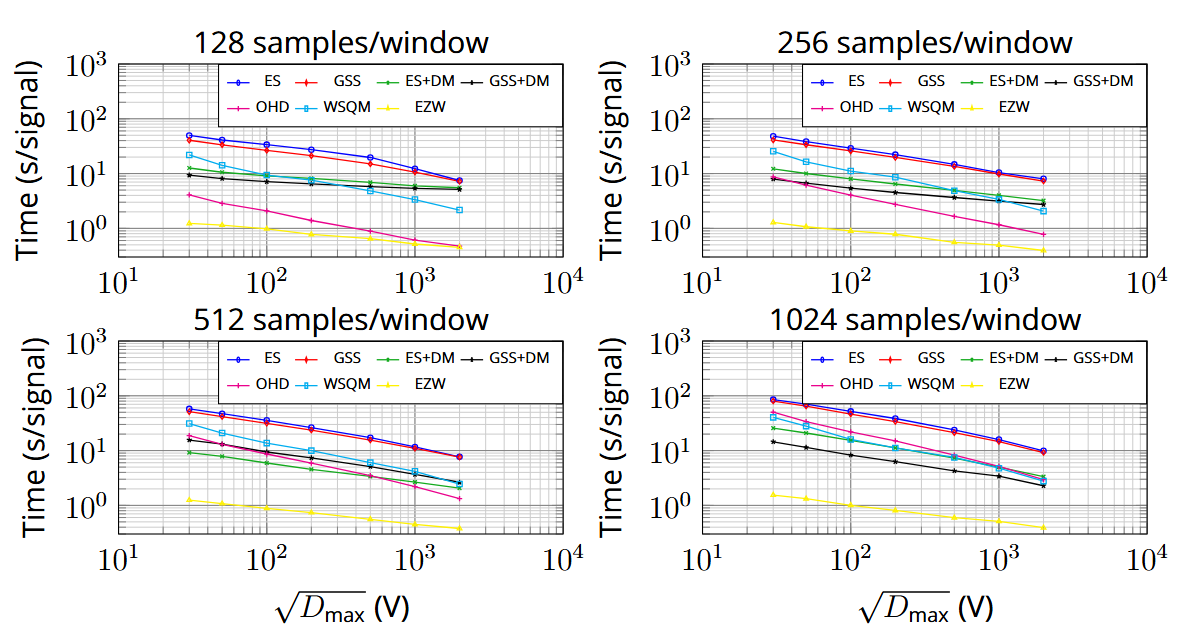}

\caption{Evolution of the average computation time required to encode each
one-second signal duration using the proposed algorithms available
at 
, for different $D_{\max}$
constraints, considering \gls{es}, \gls{es} with \gls{dm}, \gls{gss},
\gls{gss} with \gls{dm}, \gls{wsqm}, \gls{ohd}, and \gls{ezw}.
The first row corresponds to window sizes of 128 and 256 samples,
while the second row corresponds to 512 and 1024 samples, respectively.
Results are averaged over the twelve first-phase voltage signals described
in Section~\ref{subsec:Simulation set-up}.}
\label{fig:Time_R}
\end{figure}

Figure~\ref{fig:Time_R} shows that the proposed compression methods,
\gls{es} with \gls{dm} and \gls{es} with \gls{gss}, are more efficient
in terms of computational time compared to \gls{es} and \gls{gss}.
The computation time is proportional to the number of transforms performed
per window, which is detailed in Table~\ref{tab:complexity} for
each method. Compared to \gls{wsqm}, the methods \gls{es} with \gls{dm}
and \gls{es} with \gls{gss} achieve similar computation times while
providing significantly better compression performance.
Moreover, their computational advantage over \gls{wsqm} increases
with the window size $N$. However, when compared to \gls{ohd} and
\gls{ezw}, the proposed methods still require significantly more
computation time. This additional cost is the trade-off for achieving
better compression performance on average compared to existing approaches.

\begin{table}[H]
\centering
\scalebox{0.8}{
\begin{tabular}{|c|c|c|c|c|c|c|}
\hline 
\gls{es} (Sec~\ref{subsec:Exhaustive Search}) & \gls{gss} (Sec~\ref{sec:Golden Section Search})  & \gls{es} with \gls{dm} (Sec~\ref{subsec:Reduced-complexity search algorithm})  & \gls{gss} with \gls{dm} (Sec~\ref{subsec:Reduced-complexity search algorithm})  & \gls{ohd} & \gls{wsqm}  & \gls{ezw} \tabularnewline
\hline 
$\approx100O(N\log(N))$ & $\approx74O(N\log(N))$ & $\approx16O(N\log(N))$ & $\approx11O(N\log(N))$ & $\approx20O(N^{2})$ & $\approx O(N\log(N))$ & $\approx O(N\log(N))$\tabularnewline
\hline 
\end{tabular} }

\caption{Computational complexity for different methods. These values were
obtained experimentally under a constraint of $D_{\text{max}}=200^{2}\,\text{V}^{2}$.
For \gls{es} with \gls{dm}, $\Delta_{\text{M}}=3$ and $\Delta_{\text{nx}}=7$.
For \gls{ohd}, the computational cost is determined by the average
number of selected vectors ($\approx10$ for $D_{\text{max}}=200^{2}\,\text{V}^{2}$.)
multiplied by the dictionary size ($2N$) and the signal dimension
($N$).}
\label{tab:complexity} 
\end{table}

Table~\ref{tab:complexity} presents the computational complexity,
expressed as the number of transforms applied, for each method. The
results indicate that the proposed methods in this paper involve
higher computational costs than those in the literature, primarily
due to additional evaluations required for precise model selection
and bit allocation. Although these methods incur a higher computational
cost compared to reference methods, even when considering \gls{es}
with \gls{dm} or \gls{gss}, the compression performance is significantly
improved. This increase in computational cost is the trade-off required
to achieve superior performance.

\section{Conclusion}

\label{sec:Conclusion}

This paper considers the two-stage \gls{mmc} scheme for \gls{vcs}
developed in \cite{MMC} initially considering target rate constraints.
The \gls{mmc} scheme has been adapted to reach a target distortion
constraint. Three approaches have been developed for selecting the
signal model for the first stage and the optimal bit allocation between
stages in order to satisfy a target distortion constraint while minimizing
the required total rate. An \gls{es} serves as baseline.
Using \gls{gss}, the rate selection is facilitated, but the approach
may be suboptimal when the total rate required to meet the distortion
target is not a convex function of the rate required for the first
stage. The third method is based on a \gls{rd} model of the total
rate as a function of the rate of the first stage. A subset of promising
candidate models can be selected and an approximate bit allocation
for each of these models can be determined. They can then be fed to
the two previous approaches to obtained refined estimates of the best
signal model and rate allocation.

Experimental results confirm that the proposed methods outperform
established techniques, such as \gls{ohd}, \gls{wsqm}, or \gls{ezw}
in terms of rate for a given distortion constraint. Furthermore, the
reduced-complexity approaches (\gls{es} with \gls{dm} and \gls{gss})
maintain performance close to the exhaustive search but at a fraction
of the computational cost.

Future work will be dedicated to the exploitation of reinforcement
learning techniques to select the best signal model in the first stage
and the optimal bit allocation between stages accounting for the signal
characteristics and the previously selected models.

\bibliographystyle{IEEEtran}
\bibliography{References}

@article{yang2006rate,
  title={Rate control for H. 264 with two-step quantization parameter determination but single-pass encoding},
  author={Yang, Xiaokang and Tan, Yongmin and Ling, Nam},
  journal={EURASIP Journal on Advances in Signal Processing},
  volume={2006},
  number={1},
  pages={063409},
  year={2006},
  publisher={Springer}
}

@article{battista2022overview,
  title={Overview of the low complexity enhancement video coding (LCEVC) standard},
  author={Battista, Stefano and Meardi, Guido and Ferrara, Simone and Ciccarelli, Lorenzo and Maurer, Florian and Conti, Massimo and Orcioni, Simone},
  journal={IEEE Transactions on Circuits and Systems for Video Technology},
  volume={32},
  number={11},
  pages={7983--7995},
  year={2022},
  publisher={IEEE}
}

@article{zhou2020rate,
  title={Rate control method based on deep reinforcement learning for dynamic video sequences in HEVC},
  author={Zhou, Mingliang and Wei, Xuekai and Kwong, Sam and Jia, Weijia and Fang, Bin},
  journal={IEEE Transactions on Multimedia},
  volume={23},
  pages={1106--1121},
  year={2020},
  publisher={IEEE}
}

@inproceedings{huang2018qarc,
  title={QARC: Video quality aware rate control for real-time video streaming based on deep reinforcement learning},
  author={Huang, Tianchi and Zhang, Rui-Xiao and Zhou, Chao and Sun, Lifeng},
  booktitle={Proceedings of the 26th ACM international conference on Multimedia},
  pages={1208--1216},
  year={2018}
}

@article{boyce2015overview,
  title={Overview of SHVC: Scalable extensions of the high efficiency video coding standard},
  author={Boyce, Jill M and Ye, Yan and Chen, Jianle and Ramasubramonian, Adarsh K},
  journal={IEEE Transactions on Circuits and Systems for Video Technology},
  volume={26},
  number={1},
  pages={20--34},
  year={2015},
  publisher={IEEE}
}

@article{schwarz2007overview,
  title={Overview of the scalable video coding extension of the H. 264/AVC standard},
  author={Schwarz, Heiko and Marpe, Detlev and Wiegand, Thomas},
  journal={IEEE Transactions on circuits and systems for video technology},
  volume={17},
  number={9},
  pages={1103--1120},
  year={2007},
  publisher={IEEE}
}

@article{mei2021learning,
  title={Learning-based scalable image compression with latent-feature reuse and prediction},
  author={Mei, Yixin and Li, Li and Li, Zhu and Li, Fan},
  journal={IEEE Transactions on Multimedia},
  volume={24},
  pages={4143--4157},
  year={2021},
  publisher={IEEE}
}

@inproceedings{fu2024learned,
  title={Learned Image Compression with Dual-Branch Encoder and Conditional Information Coding},
  author={Fu, Haisheng and Liang, Feng and Liang, Jie and Fang, Zhenman and Zhang, Guohe and Han, Jingning},
  booktitle={2024 Data Compression Conference (DCC)},
  pages={173--182},
  year={2024},
  organization={IEEE}
}

@article{linde1980algorithm,
  title={An algorithm for vector quantizer design},
  author={Linde, Yoseph and Buzo, Andres and Gray, Robert},
  journal={IEEE Transactions on communications},
  volume={28},
  number={1},
  pages={84--95},
  year={1980},
  publisher={IEEE}
}

@standard{IEC_60044_1,
        author={{CEI IEC}},
	title = {60044-1:2003 {I}nternational {S}tandard {I}nstrument {T}ransformers},
	year={2003},
        url= {},
	pages={120},
}

@standard{IEEE_C37_118_2,
  author          = {{IEEE}},
  title        = {Standard for Synchrophasor Data Transfer for Power Systems},
  number       = {Std C37.118.2-2011},
  year         = {2011},
  address      = {Piscataway, NJ, USA},
  
}

@techreport{STTP,
  author       = {Open Group},
  title        = {Streaming Telemetry Transport Protocol {(STTP)}},
  institution  = {The Open Group},
  year         = {2021},
  url          = {https://sttp.open.org},
  note         = {Accessed: 2024-08-26}
}

@inproceedings{walker1931periodicity,
    author={Walker, Gilbert Thomas},
    title={On periodicity in series of related terms},
    booktitle={Series A, Containing Papers of a Mathematical and Physical Character},
    volume={131},
    number={818},
    pages={518--532},
    year={1931},
    publisher={The Royal Society London}
}

@online{Database_RTE, 
    author={Presv\^ots, Corentin and Prevost, Thibault},  
    title={Database of Voltage and Current Samples Values from the French Electricity Transmission Grid, Réseau de Transport d'Electricité {(RTE)}, France},
    url={https://github.com/rte-france/digital-fault-recording-database},
    year={2024},
}

@inproceedings{aklouf2021interframe,
  title={Interframe-dependent rate-QP-distortion model for video coding and transmission},
  author={Aklouf, Mourad and Leny, Marc and Kieffer, Michel and Dufaux, Fr{\'e}d{\'e}ric},
  booktitle={2021 IEEE International Conference on Image Processing (ICIP)},
  pages={2019--2023},
  year={2021},
  organization={IEEE}
}

@article{Gu2023,
    author    = {Gu, Yunjie and Green, Timothy C.},
    title     = {Power System Stability With a High Penetration of Inverter-Based Resources},
    journal   = {Proceedings of the IEEE},
    year      = {2023},
    volume    = {111},
    number    = {7},
    pages     = {832--853},
}

@article{MMC,
    author = {Presv\^ots, Corentin and Kieffer, Michel and Prevost, Thibault and Panciatici, Patrick and Zuxing, Li and Piantanida, Pablo},
    title ={Multiple-Model Coding Scheme for Electrical Signal Compression},
    journal = {Signal Processing},
    volume = {236},
    year = {2025}
}

@INPROCEEDINGS{MMC_DCC,
  author={Presv\^ots, Corentin and Kieffer, Michel and Prevost, Thibault and Panciatici, Patrick and Li, Zuxing and Piantanida, Pablo},
  booktitle={Proc. {IEEE} Data Compression Conference}, 
  title={Two-stage Multiple-Model Compression Approach for Sampled Electrical Signals}, 
  year={2024},
  volume={},
  number={},
  pages={522-531}
}

@article{nascimento2023hartley,
  title={Hartley Transform Signal Compression and Fast Power Quality Measurements for Smart Grid Application},
  author={de Olivera Nascimento, Francisco Assis},
  journal={IEEE Transactions on Power Delivery},
  year={2023},
  publisher={IEEE}
}

@inproceedings{hao2023compression,
  title={Compression of power quality disturbance using wavelet-atomic decomposition for grid-connected wind farms},
  author={Hao, Wenbo and Hu, Benran and Zhao, Leilei},
  booktitle={Proc. {IEEE} Int. Conf. Electrical Engineering and Green Energy},
  pages={171--178},
  year={2023},
}

@ARTICLE{Kapisch_Spectra_Variation_Based_Signal_Compressio2022,
  author={Kapisch, Eder B. and de Morais, Victor V. and Silva, Leandro R. M. and Filho, Luciano M. A. and Duque, Carlos A.},
  journal={IEEE Transactions on Industrial Informatics}, 
  title={Spectral Variation-Based Signal Compression Technique for Gapless Power Quality Waveform Recording in Smart Grids}, 
  year={2022},
  volume={18},
  number={7},
  pages={4488-4498},
  publisher={IEEE}
  }

@article{Ruiz_electrical_2020,
  title={Electrical faults signals restoring based on compressed sensing techniques},
  author={Ruiz, Milton and Montalvo, Iv{\'a}n},
  journal={MDPI Energies},
  volume={13},
  number={8},
  pages={2121},
  year={2020},
  publisher={MDPI}
}

@INPROCEEDINGS{Prathibha_Dual_tree_complex_wavelet2016,
  author={Prathibha, E. and Manjunatha, A. and Basavaraj, Sunil},
  booktitle={Biennial International Conference on Power and Energy Systems: Towards Sustainable Energy}, 
  title={Dual tree complex wavelet transform based approach for power quality monitoring and data compression}, 
  year={2016},
  volume={},
  number={},
  pages={1-5},
  organization={IEEE}
  }

@book{cover1999elements,
  title={Elements of information theory},
  author={Cover, Thomas M},
  year={1999},
  publisher={John Wiley \& Sons}
}

@inproceedings{nguyen2020scalable,
  title={Scalable high efficiency video coding based HTTP adaptive streaming over QUIC},
  author={Nguyen, Minh and Amirpour, Hadi and Timmerer, Christian and Hellwagner, Hermann},
  booktitle={Proceedings of the Workshop on the Evolution, Performance, and Interoperability of QUIC},
  pages={28--34},
  year={2020}
}

@inproceedings{kossentini2020svt,
  title={The SVT-AV1 encoder: overview, features and speed-quality tradeoffs},
  author={Kossentini, Faouzi and Guermazi, Hassen and Mahdi, Nader and Nouira, Chekib and Naghdinezhad, Amir and Tmar, Hassene and Khlif, Omar and Worth, Phoenix and Amara, Foued Ben},
  booktitle={Applications of Digital Image Processing XLIII},
  volume={11510},
  pages={469--490},
  year={2020},
  organization={SPIE}
}

@INPROCEEDINGS{Lorio_Analysis_of_data_compression,
  author={Lorio, F. and Magnago, F.},
  booktitle={Power Engineering Society General Meeting.}, 
  title={Analysis of data compression methods for power quality events}, 
  year={2004},
  volume={1},
  number={},
  pages={504-509},
  organization={IEEE}
  }

@inproceedings{klump_lossless_2010,
	  author={Klump, Ray and Agarwal, Pooja and Tate, Joseph Euzebe and Khurana, Himanshu},
	  booktitle={Power Engineering SocietyGeneral Meeting}, 
	  title={Lossless compression of synchronized phasor measurements}, 
	  year={2010},
	  volume={},
	  number={},
	  pages={1-7},
        organization={IEEE}
	  }

@article{hamid_wavelet-based_2002,
	  author={Hamid, E.Y. and Kawasaki, Z.-I.},
	  journal={IEEE Transactions on Power Delivery}, 
	  title={Wavelet-based data compression of power system disturbances using the minimum description length criterion}, 
	  year={2002},
	  volume={17},
	  number={2},
	  pages={460-466},
        publisher={IEEE}
	  }

@article{ning_wavelet-based_2011,
	  author={Ning, Jiaxin and Wang, Jianhui and Gao, Wenzhong and Liu, Cong},
	  journal={IEEE Transactions on Smart Grid}, 
	  title={A Wavelet-Based Data Compression Technique for Smart Grid}, 
	  year={2011},
	  volume={2},
	  number={1},
	  pages={212-218},
        publisher={IEEE}
	  }

@article{khan_embedded-zerotree-wavelet-based_2015,
	  author={Khan, Jesmin and Bhuiyan, Sharif M. A. and Murphy, Gregory and Arline, Morgan},
	  journal={IEEE Transactions on Industry Applications}, 
	  title={Embedded-Zerotree-Wavelet-Based Data Denoising and Compression for Smart Grid}, 
	  year={2015},
	  volume={51},
	  number={5},
	  pages={4190-4200},
        publisher={IEEE}
	  }

@article{littler_wavelets_1999,
  author={Littler, T.B. and Morrow, D.J.},
  journal={IEEE Transactions on Power Delivery}, 
  title={Wavelets for the analysis and compression of power system disturbances}, 
  year={1999},
  volume={14},
  number={2},
  pages={358-364},
  publisher={IEEE}
  }

@article{tse_real-time_2012,  
author={Tse, Norman C. F. and Chan, John Y. C. and Lau, Wing-Hong and Poon, Jone T. Y. and Lai, L. L.},
journal={IEEE Transactions on Power Delivery}, 
title={Real-Time Power-Quality Monitoring With Hybrid Sinusoidal and Lifting Wavelet Compression Algorithm}, 
year={2012},
volume={27},
number={4},
pages={1718-1726},
publisher={IEEE}
}

@article{cormane_spectral_2016,
author={Cormane, Jorge and de O. Nascimento, Francisco Assis},
journal={IEEE Transactions on Smart Grid}, 
title={Spectral Shape Estimation in Data Compression for Smart Grid Monitoring}, 
year={2016},
volume={7},
number={3},
pages={1214-1221},
publisher={IEEE}
}

@article{nascimento_improved_2020,
author={Nascimento, Francisco A. de O. and Saraiva, Raimundo G. and Cormane, Jorge},
journal={IEEE Transactions on Power Delivery}, 
title={Improved Transient Data Compression Algorithm Based on Wavelet Spectral Quantization Models}, 
year={2020},
volume={35},
number={5},
pages={2222-2232},
publisher={IEEE}
}

@ARTICLE{Lovisolo_Efficient_coherent_adaptive_representations2005,
  author={Lovisolo, L. and da Silva, E.A.B. and Rodrigues, M.A.M. and Diniz, P.S.R.},
  journal={IEEE Trans. Signal Processing}, 
  title={Efficient coherent adaptive representations of monitored electric signals in power systems using damped sinusoids}, 
  year={2005},
  volume={53},
  number={10},
  pages={3831-3846},
  publisher={IEEE}
  }

@article{tcheou_optimum_2007,
  author={Tcheou, Michel P. and Lovisolo, Lisandro and da Silva, Eduardo A. B. and Rodrigues, Marco A. M. and Diniz, Paulo S. R.},
  journal={IEEE Signal Processing Letters}, 
  title={Optimum Rate-Distortion Dictionary Selection for Compression of Atomic Decompositions of Electric Disturbance Signals}, 
  year={2007},
  volume={14},
  number={2},
  pages={81-84},
  publisher={IEEE}
  }

@article{sabarimalai_manikandan_simultaneous_2015,
  title={Simultaneous denoising and compression of power system disturbances using sparse representation on overcomplete hybrid dictionaries},
  author={Sabarimalai Manikandan, M and Samantaray, Subhransu Ranjan and Kamwa, Innocent},
  journal={Wiley Online Library IET Generation, Transmission \& Distribution},
  volume={9},
  number={11},
  pages={1077--1088},
  year={2015},
  publisher={Wiley Online Library}
}

@article{meher_integrated_2004,
  title={An integrated data compression scheme for power quality events using spline wavelet and neural network},
  author={Meher, Saroj K and Pradhan, AK and Panda, G},
  journal={Elsevier Electric Power Systems Research},
  volume={69},
  number={2-3},
  pages={213--220},
  year={2004},
  publisher={Elsevier}
}

@article{wang_adaptive_2021,
  author={Wang, Xinyi and Liu, Yilu and Tong, Lang},
  journal={IEEE Transactions on Power Systems}, 
  title={Adaptive Subband Compression for Streaming of Continuous Point-on-Wave and {PMU} Data}, 
  year={2021},
  volume={36},
  number={6},
  pages={5612-5621},
  organization={IEEE}
  }

@article{ribeiro_novel_2007,
  author={Ribeiro, Moiss V. and Park, Seop Hyeong and Romano, Joo Marcos T. and Mitra, Sanjit K.},
  journal={IEEE Transactions on Power Delivery}, 
  title={A Novel MDL-based Compression Method for Power Quality Applications}, 
  year={2007},
  volume={22},
  number={1},
  pages={27-36},
  organization={IEEE}
  }

@article{zygarlicki_data_2006,
  title={Data compression using Prony's method and wavelet transform in power quality monitoring systems},
  author={Zygarlicki, Jaroslaw and Mroczka, Janusz},
  journal={PAN Metrology and Measurement Systems},
  volume={13},
  number={3},
  pages={325--251},
  year={2006},
  publisher={PAN}
}

@article{lovisolo_modeling_2007,
  title={Modeling of electric disturbance signals using damped sinusoids via atomic decompositions and its applications},
  author={Lovisolo, Lisandro and Tcheou, Michel P and da Silva, Eduardo AB and Rodrigues, Marco AM and Diniz, Paulo SR},
  journal={{EURASIP} Journal on Advances in Signal Processing},
  volume={2007},
  pages={1--15},
  year={2007},
  publisher={Springer}
}

@article{he_high_2019,
  author={He, Shunfan and Tian, Wei and Zhang, Junmin and Li, Kaicheng and Zhang, Ming and Zhu, Rongbo},
  journal={IEEE Transactions on Industrial Informatics}, 
  title={A High Efficient Approach for Power Disturbance Waveform Compression in the View of Heisenberg Uncertainty}, 
  year={2019},
  volume={15},
  number={5},
  pages={2580-2591},
  publisher={IEEE}
  }

@inproceedings{de_oliveira_artificial_2018,
  author={de Oliveira, Gabriel A. and Tcheou, Michel P. and Lovisolo, Lisandro},
  booktitle={International Joint Conference on Neural Networks}, 
  title={Artificial Neural Networks For Dictionary Selection in Adaptive Greedy Decomposition Algorithms With Reduced Complexity}, 
  year={2018},
  volume={},
  number={},
  pages={1-8},
  organization={IEEE}
  }

@article{zhang_high_2011,
  author={Zhang, Ming and Li, Kaicheng and Hu, Yisheng},
  journal={IEEE Transactions on Instrumentation and Measurement}, 
  title={A High Efficient Compression Method for Power Quality Applications}, 
  year={2011},
  volume={60},
  number={6},
  pages={1976--1985},
  publisher={IEEE}
  }

@inproceedings{qing_compression_2011,
  author={Qing, An and Hongtao, Zhang and Zhikun, Hu and Zhiwen, Chen},
  booktitle={Proc. {IEEE} Int. Conf. Measuring Technology and Mechatronics Automation}, 
  title={A Compression Approach of Power Quality Monitoring Data Based on Two-dimension {DCT}}, 
  year={2011},
  volume={1},
  number={},
  pages={20-24}
  }

@inproceedings{dapper_high_2015,
  author={Dapper, Roque Eduardo and Susin, Altamiro Amadeu and Bampi, Sergio and Crovato, César David Paredes},
  booktitle={International Symposium on Industrial Electronics}, 
  title={High compression ratio algorithm for power quality signals}, 
  year={2015},
  volume={},
  number={},
  pages={1322-1326},
  organization={IEEE},
  }

@article{khan_weighted_2021,
  title={Weighted entropy and modified MDL for compression and denoising data in smart grid},
  author={Khan, Jesmin},
  journal={Elsevier International Journal of Electrical Power \& Energy Systems},
  volume={133},
  pages={107089},
  year={2021},
  publisher={Elsevier}
}

@article{zygarlicki_reduced_2010,
  author={Zygarlicki, Jaroslaw and Zygarlicka, Malgorzata and Mroczka, Janusz and Latawiec, Krzysztof J.},
  journal={IEEE Transactions on Power Delivery}, 
  title={A Reduced Prony's Method in Power-Quality Analysis-Parameters Selection}, 
  year={2010},
  volume={25},
  number={2},
  pages={979-986},
  publisher={IEEE}
  }

@article{shapiro_embedded_1993,
  author={Shapiro, J.M.},
  journal={IEEE Transactions on Signal Processing}, 
  title={Embedded image coding using zerotrees of wavelet coefficients}, 
  year={1993},
  volume={41},
  number={12},
  pages={3445-3462},
  publisher={IEEE}
  }

@article{balle_variational_2018,
  title={Variational image compression with a scale hyperprior},
  author={Ball{\'e}, Johannes and Minnen, David and Singh, Saurabh and Hwang, Sung Jin and Johnston, Nick},
  journal={arXiv preprint arXiv:1802.01436},
  year={2018}
}

@article{panter1951quantization,
  title={Quantization distortion in pulse-count modulation with nonuniform spacing of levels},
  author={Panter, PF and Dite, Wu},
  journal={Proceedings of the IRE},
  volume={39},
  number={1},
  pages={44--48},
  year={1951},
  publisher={IEEE}
}

@article{balle_end--end_2017,
  title={End-to-end optimized image compression},
  author={Ball{\'e}, Johannes and Laparra, Valero and Simoncelli, Eero P},
  journal={arXiv preprint arXiv:1611.01704},
  year={2016}
}

@article{lee_context-adaptive_2019,
  title={Context-adaptive entropy model for end-to-end optimized image compression},
  author={Lee, Jooyoung and Cho, Seunghyun and Beack, Seung-Kwon},
  journal={arXiv preprint arXiv:1809.10452},
  year={2018}
}

@article{Kamps_chebyshev_polynomials_1989,
  title={Chebyshev polynomials and least squares estimation based on one-dependent random variables},
  author={Kamps, Udo},
  journal={Elsevier Linear Algebra and its Applications},
  volume={112},
  pages={217--230},
  year={1989},
  publisher={Elsevier}
}

@book{sayood_introduction_2017,
  author       = {Khalid Sayood},
  title        = {Introduction to Data Compression},
  year         = {2017},
  edition      = {5th},
  publisher    = {Morgan Kaufmann},
  address      = {San Francisco, CA},
  chapter      = {Fixed-Rate Compression},
  isbn         = {978-0128094747}
}

@inproceedings{hoang_new_2009,
  author={Hoang, Thi Minh Nguyet and Ragot, Stephane and Oger, Marie and Antonini, Marc},
  booktitle={International Conference on Acoustics, Speech and Signal Processing}, 
  title={A new bitplane coder for scalable transform audio coding}, 
  year={2009},
  volume={},
  number={},
  pages={4137-4140},
  organization={IEEE}
  }

@inproceedings{chung_variable_1999,
  author={Jaehak Chung and Powers, E.J. and Grady, W.M. and Bhatt, S.C.},
  booktitle={Power Engineering Society}, 
  title={Variable rate power disturbance signal compression using embedded zerotree wavelet transform coding}, 
  year={1999},
  volume={2},
  number={},
  pages={1305-1309 vol.2},
  organization={IEEE},
  }

@article{kong2023mixture,
  title={Mixture autoregressive and spectral attention network for multispectral image compression based on variational autoencoder},
  author={Kong, Fanqiang and Ren, Guanglong and Hu, Yunfang and Li, Dan and Hu, Kedi},
  journal={The Visual Computer},
  pages={1--24},
  year={2023},
  publisher={Springer}
}

@article{zhang2023exploring,
  title={Exploring resolution fields for scalable image compression with uncertainty guidance},
  author={Zhang, Dongyi and Li, Feng and Liu, Man and Cong, Runmin and Bai, Huihui and Wang, Meng and Zhao, Yao},
  journal={IEEE Transactions on Circuits and Systems for Video Technology},
  year={2023},
  publisher={IEEE}
}

@article{santoso1997power,
  title={Power quality disturbance data compression using wavelet transform methods},
  author={Santoso, Surya and Powers, Edward J and Grady, W Mack},
  journal={IEEE Transactions on Power Delivery},
  volume={12},
  number={3},
  pages={1250--1257},
  year={1997},
  publisher={IEEE}
}

@article{gu2003bridge,
  title={Bridge the gap: signal processing for power quality applications},
  author={Gu, Irene Yu-Hua and Styvaktakis, Emmanouil},
  journal={Elsevier Electric Power Systems Research},
  volume={66},
  number={1},
  pages={83--96},
  year={2003},
  publisher={Elsevier}
}

@article{bollen2009bridging,
  title={Bridging the gap between signal and power},
  author={Bollen, Math HJ and Gu, Irene YH and Santoso, Surya and McGranaghan, Mark F and Crossley, Peter A and Ribeiro, Moises V and Ribeiro, Paulo F},
  journal={IEEE Signal processing magazine},
  volume={26},
  number={4},
  pages={12--31},
  year={2009},
  publisher={IEEE}
}

@article{khokhar2017new,
  title={A new optimal feature selection algorithm for classification of power quality disturbances using discrete wavelet transform and probabilistic neural network},
  author={Khokhar, Suhail and Zin, Abdullah Asuhaimi Mohd and Memon, Aslam Pervez and Mokhtar, Ahmad Safawi},
  journal={Elsevier Measurement},
  volume={95},
  pages={246--259},
  year={2017},
  publisher={Elsevier}
}

@article{yan2023review,
	author = {Yanjun Yan and Kai Chen and Hang Geng and Wenqian Fan and Xinrui Zhou},
	title = {A Review on Intelligent Detection and Classification of Power Quality Disturbances: Trends, Methodologies, and Prospects},
	journal = {Computer Modeling in Engineering and Sciences},
	volume = {137},
	number = {2},
	pages = {1345-1379},
	year = {2023}
}

@article{panda2002data,
  title={Data compression of power quality events using the slantlet transform},
  author={Panda, Ganapati and Dash, PK and Pradhan, Ashok Kumar and Meher, Saroj K},
  journal={IEEE Transactions on power delivery},
  volume={17},
  number={2},
  pages={662--667},
  year={2002},
  publisher={IEEE}
}

@article{hsieh2003disturbance,
  title={Disturbance data compression of a power system using the Huffman coding approach with wavelet transform enhancement},
  author={Hsieh, C-T and Huang, S-J},
  journal={IET Proceedings: Generation, Transmission and Distribution},
  volume={150},
  number={1},
  pages={7--14},
  year={2003},
  publisher={IET}
}

@article{dash2003power,
  title={Power quality disturbance data compression, detection, and classification using integrated spline wavelet and S-transform},
  author={Dash, PK and Panigrahi, BK and Sahoo, DK and Panda, G},
  journal={IEEE Transactions on Power Delivery},
  volume={18},
  number={2},
  pages={595--600},
  year={2003},
  publisher={IEEE}
}

@inproceedings{wu2003data,
  title={Data compression technique in recording electric arc furnace voltage and current waveforms for tracking power quality},
  author={Wu, Chi-Jui and Fu, Tsu-Hsun and Huang, Cheng-Ping},
  booktitle={Proc. {IEEE} Transmission and Distribution Conference and Exposition},
  volume={1},
  pages={383--388},
  year={2003},
  organization={IEEE}
}

@inproceedings{shang2003efficiency,
  title={Efficiency analysis of data compression of power system transients using wavelet transform},
  author={Shang, L and Jaeger, J and Krebs, R},
  booktitle={Bologna Power Tech Conference Proceedings},
  volume={4},
  pages={6--pp},
  year={2003},
  organization={IEEE}
}

@article{huang2004application,
  title={Application of arithmetic coding for electric power disturbance data compression with wavelet packet enhancement},
  author={Huang, Shyh-Jier and Jou, Ming-Jong},
  journal={IEEE Transactions on Power Systems},
  volume={19},
  number={3},
  pages={1334--1341},
  year={2004},
  publisher={IEEE}
}

@inproceedings{yuan2006power,
  title={Power system fault data compression using the wavelet transform and vector quantification},
  author={Yuan, Yue and Yu, Xiaoming and Du, Hongji},
  booktitle={International Conference on Power System Technology},
  pages={1--6},
  year={2006},
  organization={IEEE}
}

@article{rissanen1978modeling,
  title={Modeling by shortest data description},
  author={Rissanen, Jorma},
  journal={Elsevier Automatica},
  volume={14},
  number={5},
  pages={465--471},
  year={1978},
  publisher={Elsevier}
}

@article{ribeiro2007classification,
  title={Classification of single and multiple disturbances in electric signals},
  author={Ribeiro, Mois{\'e}s Vidal and Pereira, Jos{\'e} Luiz Rezende},
  journal={{EURASIP} Journal on Advances in Signal Processing},
  volume={2007},
  pages={1--18},
  year={2007},
  publisher={Springer}
}

@inproceedings{chung2000electric,
  title={Electric power transient disturbance classification using wavelet-based hidden Markov models},
  author={Chung, Jaehak and Powers, Edward J and Grady, W Mack and Bhatt, Sid C},
  booktitle={International Conference on Acoustics, Speech, and Signal Processing},
  volume={6},
  pages={3662--3665},
  year={2000},
  organization={IEEE}
}

@inproceedings{de1997data,
  title={Data compression algorithm for transient recording system},
  author={de Oliveira Nascimento, F Assis},
  booktitle={International Symposium on Industrial Electronics},
  pages={1126--1130},
  year={1997},
  organization={IEEE}
}

@article{mallat1993matching,
  title={Matching pursuits with time-frequency dictionaries},
  author={Mallat, St{\'e}phane G and Zhang, Zhifeng},
  journal={IEEE Transactions on signal processing},
  volume={41},
  number={12},
  pages={3397--3415},
  year={1993},
  publisher={IEEE}
}

@article{tcheou2012far,
  title={How far can one compress digital fault records? Analysis of a matching pursuit-based algorithm},
  author={Tcheou, Michel P and Miranda, Andr{\'e} LL and Lovisolo, Lisandro and da Silva, Eduardo AB and Rodrigues, Marco AM and Diniz, Paulo SR},
  journal={Elsevier Digital Signal Processing},
  volume={22},
  number={2},
  pages={288--297},
  year={2012},
  publisher={Elsevier}
}

@book{aggarwal2023neural,
  title={Neural Networks and Deep Learning: A Textbook},
  author={Aggarwal, Charu C.},
  year={2023},
  publisher={Springer}
}

@article{paolone_fundamentals_2020,
  title={Fundamentals of power systems modelling in the presence of converter-interfaced generation},
  author={Paolone, Mario and Gaunt, Trevor and Guillaud, Xavier and Liserre, Marco and Meliopoulos, Sakis and Monti, Antonello and Van Cutsem, Thierry and Vittal, Vijay and Vournas, Costas},
  journal={Electric Power Systems Research},
  volume={189},
  pages={106811},
  year={2020},
  publisher={Elsevier}
}

@article{karpilow_characterization_2021,
  title={Characterization of non-stationary signals in electric grids: A functional dictionary approach},
  author={Karpilow, Alexandra and Dervi{\v{s}}kadi{\'c}, Asja and Frigo, Guglielmo and Paolone, Mario},
  journal={IEEE Transactions on Power Systems},
  volume={37},
  number={2},
  pages={1126--1138},
  year={2021},
  publisher={IEEE}
}

@techreport{karpilow2024functional,
  title={Functional-Basis Analysis of Non-Stationary Signals in Modern Power Grids: Theory and Implementation in Embedded Systems},
  author={Karpilow, Alexandra Cameron},
  year={2024},
  institution={EPFL}
}

@article{yaghoobi2009parametric,
  title={Parametric dictionary design for sparse coding},
  author={Yaghoobi, Mehrdad and Daudet, Laurent and Davies, Mike E},
  journal={IEEE Transactions on Signal Processing},
  volume={57},
  number={12},
  pages={4800--4810},
  year={2009},
  publisher={IEEE}
}

@article{alves2021reduced,
  title={Reduced-complexity end-to-end variational autoencoder for on board satellite image compression},
  author={Alves de Oliveira, Vinicius and Chabert, Marie and Oberlin, Thomas and Poulliat, Charly and Bruno, Mickael and Latry, Christophe and Carlavan, Mikael and Henrot, Simon and Falzon, Frederic and Camarero, Roberto},
  journal={Remote Sensing},
  volume={13},
  number={3},
  pages={447},
  year={2021},
  publisher={MDPI}
}

@article{ding1996rate,
  title={Rate control of MPEG video coding and recording by rate-quantization modeling},
  author={Ding, Wei and Liu, Bede},
  journal={IEEE transactions on circuits and systems for video technology},
  volume={6},
  number={1},
  pages={12--20},
  year={1996},
  publisher={IEEE}
}

@article{choi2013pixel,
  title={Pixel-wise unified rate-quantization model for multi-level rate control},
  author={Choi, Hyomin and Yoo, Jonghun and Nam, Junghak and Sim, Donggyu and Baji{\'c}, Ivan V},
  journal={IEEE Journal of Selected Topics in Signal Processing},
  volume={7},
  number={6},
  pages={1112--1123},
  year={2013},
  publisher={IEEE}
}

@article{ma2005rate,
  title={Rate-distortion analysis for H. 264/AVC video coding and its application to rate control},
  author={Ma, Siwei and Gao, Wen and Lu, Yan},
  journal={IEEE transactions on circuits and systems for video technology},
  volume={15},
  number={12},
  pages={1533--1544},
  year={2005},
  publisher={IEEE}
}

@inproceedings{li2014inter,
  title={Inter-dependent rate-distortion modeling for video coding and its application to rate control},
  author={Li, Yuan and Jia, Huizhu and Ma, Pan and Zhu, Chuang and Xie, Xiaodong and Gao, Wen},
  booktitle={2014 IEEE International Conference on Multimedia and Expo (ICME)},
  pages={1--6},
  year={2014},
  organization={IEEE}
}

@inproceedings{liu2010low,
  title={Low-complexity rate control based on $\rho$-domain model for scalable video coding},
  author={Liu, Meng and Guo, Yi and Li, Houqiang and Chen, Chang Wen},
  booktitle={2010 IEEE International Conference on Image Processing},
  pages={1277--1280},
  year={2010},
  organization={IEEE}
}

@article{tang2019generalized,
  title={A Generalized Rate-Distortion-$\lambda$ Model Based HEVC Rate Control Algorithm},
  author={Tang, Minhao and Wen, Jiangtao and Han, Yuxing},
  journal={arXiv preprint arXiv:1911.00639},
  year={2019}
}

@article{li2014lambda,
  title={$\lambda$ domain rate control algorithm for High Efficiency Video Coding},
  author={Li, Bin and Li, Houqiang and Li, Li and Zhang, Jinlei},
  journal={IEEE transactions on Image Processing},
  volume={23},
  number={9},
  pages={3841--3854},
  year={2014},
  publisher={IEEE}
}

@article{li2016lambda,
  title={$\lambda$-domain optimal bit allocation algorithm for High Efficiency Video Coding},
  author={Li, Li and Li, Bin and Li, Houqiang and Chen, Chang Wen},
  journal={IEEE Transactions on Circuits and Systems for Video Technology},
  volume={28},
  number={1},
  pages={130--142},
  year={2016},
  publisher={IEEE}
}

@article{lin1998bit,
  title={Bit-rate control using piecewise approximated rate-distortion characteristics},
  author={Lin, Liang-Jin and Ortega, Antonio},
  journal={IEEE Transactions on Circuits and Systems for Video Technology},
  volume={8},
  number={4},
  pages={446--459},
  year={1998},
  publisher={IEEE}
}

@article{zhang2003constant,
  title={Constant quality constrained rate allocation for FGS-coded video},
  author={Zhang, Xi Min and Vetro, Anthony and Shi, Yun Q and Sun, Huifang},
  journal={IEEE Transactions on Circuits and Systems for Video Technology},
  volume={13},
  number={2},
  pages={121--130},
  year={2003},
  publisher={IEEE}
}

@article{zhao2015ssim,
  title={SSIM-based coarse-grain scalable video coding},
  author={Zhao, Tiesong and Wang, Jiheng and Wang, Zhou and Chen, Chang Wen},
  journal={IEEE Transactions on Broadcasting},
  volume={61},
  number={2},
  pages={210--221},
  year={2015},
  publisher={IEEE}
}

@article{gupta2012h,
  title={H. 264 coarse grain scalable (CGS) and medium grain scalable (MGS) encoded video: A trace based traffic and quality evaluation},
  author={Gupta, Rohan and Pulipaka, Akshay and Seeling, Patrick and Karam, Lina J and Reisslein, Martin},
  journal={IEEE Transactions on Broadcasting},
  volume={58},
  number={3},
  pages={428--439},
  year={2012},
  publisher={IEEE}
}

@article{skodras2001jpeg,
  title={The JPEG 2000 still image compression standard},
  author={Skodras, Athanassios and Christopoulos, Charilaos and Ebrahimi, Touradj},
  journal={IEEE Signal processing magazine},
  volume={18},
  number={5},
  pages={36--58},
  year={2001},
  publisher={IEEE}
}

@article{taubman1994multirate,
  title={Multirate 3-D subband coding of video},
  author={Taubman, David and Zakhor, Avideh},
  journal={IEEE Transactions on image processing},
  volume={3},
  number={5},
  pages={572--588},
  year={1994},
  publisher={IEEE}
}

@article{said1996new,
  title={A new, fast, and efficient image codec based on set partitioning in hierarchical trees},
  author={Said, Amir and Pearlman, William A},
  journal={IEEE Transactions on circuits and systems for video technology},
  volume={6},
  number={3},
  pages={243--250},
  year={1996},
  publisher={IEEE}
}

@article{taubman2000high,
  title={High performance scalable image compression with EBCOT},
  author={Taubman, David},
  journal={IEEE Transactions on image processing},
  volume={9},
  number={7},
  pages={1158--1170},
  year={2000},
  publisher={IEEE}
}

@InBook{PressC10:2011,
  author    = {Press, W. H. and Teukolsky, S. A. and Vetterling, W. T. and Flannery, B. P.},
  booktitle = {Numerical Recipes: The Art of Scientific Computing (3rd ed.)},
  date      = {2011},
  title     = {Golden Section Search in One Dimension},
  chapter   = {10.2},
  publisher = {Cambridge University Press},
}

\appendix

\subsection{Proof of Propositions~\ref{prop:model_eq} and \ref{prop:model_eq2}}

\label{subsec:Proof of proposition model eq2} The first-order Taylor
expansion of $\boldsymbol{x}^{m}\left(\widehat{\boldsymbol{\theta}}\right)$
at $\widehat{\boldsymbol{\theta}}_{\mathrm{q}}\left(n_{\text{x}}\right)$
is
\begin{equation}
\boldsymbol{x}^{m}\left(\widehat{\boldsymbol{\theta}}\right)\simeq\boldsymbol{x}^{m}\left(\widehat{\boldsymbol{\theta}}_{\mathrm{q}}\left(n_{\text{x}}\right)\right)+\frac{\partial\boldsymbol{x}^{m}\left(\widehat{\boldsymbol{\theta}}\right)}{\partial\widehat{\boldsymbol{\theta}}^{T}}\boldsymbol{\varepsilon}\left(\widehat{\boldsymbol{\theta}},n_{\text{x}}\right),
\end{equation}
where $\boldsymbol{\varepsilon}\left(\widehat{\boldsymbol{\theta}},n_{\text{x}}\right)=\widehat{\boldsymbol{\theta}}-\widehat{\boldsymbol{\theta}}_{\mathrm{q}}\left(n_{\text{x}}\right)$.
Consequently, 
\begin{equation}
\boldsymbol{e}_{\text{q}}^{m}\left(\widehat{\boldsymbol{\theta}},n_{\mathrm{x}}\right)=\frac{\partial\boldsymbol{x}^{m}\left(\widehat{\boldsymbol{\theta}}\right)}{\partial\widehat{\boldsymbol{\theta}}^{T}}\boldsymbol{\varepsilon}\left(\widehat{\boldsymbol{\theta}},n_{\text{x}}\right).
\end{equation}
The biased estimate of the autocorrelation function of $\boldsymbol{e}_{\text{q}}^{m}\left(\widehat{\boldsymbol{\theta}},n_{\mathrm{x}}\right)$
is
\begin{align}
 & \widehat{\gamma}_{\text{q}}^{m}\left(p,\widehat{\boldsymbol{\theta}},n_{\mathrm{x}}\right)=\frac{1}{N}\sum_{n=p+1}^{N}e_{\text{q},n}^{m}\left(\widehat{\boldsymbol{\theta}},n_{\mathrm{x}}\right)e_{\text{q},n-p}^{m}\left(\widehat{\boldsymbol{\theta}},n_{\mathrm{x}}\right)\\
 & =\frac{1}{N}\text{Tr}\left(\left(\begin{array}{ccc}
\frac{\partial x_{p+1}^{m}\left(\widehat{\boldsymbol{\theta}}\right)}{\partial\widehat{\theta}_{1}} & \cdots & \frac{\partial x_{p+1}^{m}\left(\widehat{\boldsymbol{\theta}}\right)}{\partial\widehat{\theta}_{K}}\\
\vdots & \ddots & \vdots\\
\frac{\partial x_{N}^{m}\left(\widehat{\boldsymbol{\theta}}\right)}{\partial\widehat{\theta}_{1}} & \cdots & \frac{\partial x_{N}^{m}\left(\widehat{\boldsymbol{\theta}}\right)}{\partial\widehat{\theta}_{K}}
\end{array}\right)\left(\begin{array}{ccc}
\varepsilon_{1}^{2} & \cdots & \varepsilon_{1}\varepsilon_{K}\\
\vdots & \ddots & \vdots\\
\varepsilon_{K}\varepsilon_{1} & \cdots & \varepsilon_{K}^{2}
\end{array}\right)\left(\begin{array}{ccc}
\frac{\partial x_{1}^{m}\left(\widehat{\boldsymbol{\theta}}\right)}{\partial\widehat{\theta}_{1}} & \cdots & \frac{\partial x_{N-p}^{m}\left(\widehat{\boldsymbol{\theta}}\right)}{\partial\widehat{\theta}_{1}}\\
\vdots & \ddots & \vdots\\
\frac{\partial x_{1}^{m}\left(\widehat{\boldsymbol{\theta}}\right)}{\partial\widehat{\theta}_{K}} & \cdots & \frac{\partial x_{N-p}^{m}\left(\widehat{\boldsymbol{\theta}}\right)}{\partial\widehat{\theta}_{K}}
\end{array}\right)\right)\\
 & =\frac{1}{N}\text{Tr}\left(\left(\begin{array}{ccc}
\sum_{k=1}^{K}\sum_{\ell=1}^{K}\varepsilon_{k}\varepsilon_{\ell}\frac{\partial x_{p+1}^{m}\left(\widehat{\boldsymbol{\theta}}\right)}{\partial\widehat{\theta}_{k}}\frac{\partial x_{1}^{m}\left(\widehat{\boldsymbol{\theta}}\right)}{\partial\widehat{\theta}_{\ell}} & \cdots & *\\
\vdots & \ddots & \vdots\\*
* & \cdots & \sum_{k=1}^{K}\sum_{\ell=1}^{K}\varepsilon_{k}\varepsilon_{\ell}\frac{\partial x_{N}^{m}\left(\widehat{\boldsymbol{\theta}}\right)}{\partial\widehat{\theta}_{k}}\frac{\partial x_{N-p}^{m}\left(\widehat{\boldsymbol{\theta}}\right)}{\partial\widehat{\theta}_{\ell}}
\end{array}\right)\right).
\end{align}
Taking the expectation with respect to $\boldsymbol{\varepsilon}\left(\widehat{\boldsymbol{\theta}},n_{\text{x}}\right)$,
one obtains 
\begin{equation}
\mathbb{E}_{\boldsymbol{\varepsilon}}\left[\widehat{\gamma}_{\text{q}}^{m}\left(p,\widehat{\boldsymbol{\theta}},n_{\mathrm{x}}\right)\right]=\mathbb{E}_{\boldsymbol{\varepsilon}}\left[\frac{1}{N}\sum_{n=p+1}^{N}\sum_{k=1}^{K}\sum_{\ell=1}^{K}\varepsilon_{k}\varepsilon_{\ell}\frac{\partial x_{n}^{m}\left(\widehat{\boldsymbol{\theta}}\right)}{\partial\widehat{\theta}_{k}}\frac{\partial x_{n-p}^{m}\left(\widehat{\boldsymbol{\theta}}\right)}{\partial\widehat{\theta}_{\ell}}\right].
\end{equation}
The components $\varepsilon_{k}\left(\widehat{\boldsymbol{\theta}},n_{\text{x}}\right)$
of $\boldsymbol{\varepsilon}\left(\widehat{\boldsymbol{\theta}},n_{\text{x}}\right)$
are assumed zero-mean and uncorrelated for all $k=1,\dots,K$. Consequently
\begin{align}
\widehat{\gamma}_{\text{q}}^{m}\left(p,\widehat{\boldsymbol{\theta}},n_{\mathrm{x}}\right) & =\mathbb{E}_{\boldsymbol{\varepsilon}}\left[\widehat{\gamma}_{\text{q}}^{m}\left(p,\widehat{\boldsymbol{\theta}},n_{\mathrm{x}}\right)\right]\\
 & =\sum_{k=1}^{K}h_{k}\left(p,\widehat{\boldsymbol{\theta}}\right)\mathbb{E}_{\varepsilon}\left(\boldsymbol{\varepsilon}_{k}^{2}\left(\widehat{\boldsymbol{\theta}},n_{\text{x}}\right)\right),
\end{align}
with
\begin{equation}
h_{k}\left(p,\widehat{\boldsymbol{\theta}}\right)=\frac{1}{N}\sum_{n=p+1}^{N}\frac{\partial x_{n}^{m}\left(\widehat{\boldsymbol{\theta}}\right)}{\partial\widehat{\theta}_{k}}\frac{\partial x_{n-p}^{m}\left(\widehat{\boldsymbol{\theta}}\right)}{\partial\widehat{\theta}_{k}}.
\label{eq:hkp}
\end{equation}
The result of Proposition~\ref{prop:model_eq} is obtained taking $p=0$ in \eqref{eq:hkp}.

\subsection{Auto-correlation function for sinusoidal model\label{subsec:gamma_sin}}

Consider the sinusoidal model with a single sinusoidal component \eqref{eq:sin}
with estimated parameters $\widehat{\boldsymbol{\theta}}=\left(\widehat{a},\widehat{f},\widehat{\phi}\right)$.
Using Proposition~\ref{prop:model_eq2}, one has to evaluate
\begin{align*}
\frac{\partial x_{n}^{m}\left(\widehat{\boldsymbol{\theta}}\right)}{\partial\widehat{a}} & =\cos\left(2\pi\widehat{f}nT_{\text{s}}+\widehat{\phi}\right),
\end{align*}
which leads to 
\begin{align*}
h_{1}\left(p,\widehat{\boldsymbol{\theta}}\right) & =\frac{1}{N}\sum_{n=p+1}^{N}\cos\left(2\pi\widehat{f}nT_{\text{s}}+\widehat{\phi}\right)\cos\left(2\pi\widehat{f}\left(n-p\right)T_{\text{s}}+\widehat{\phi}\right)\\
 & =\frac{1}{2N}\sum_{n=p+1}^{N}\left(\cos\left(2\pi\widehat{f}\left(2n-p\right)T_{\text{s}}+2\widehat{\phi}\right)+\cos\left(2\pi\widehat{f}pT_{\text{s}}\right)\right).
\end{align*}
Then 
\begin{align*}
\sum_{n=p+1}^{N}\cos\left(2\pi\widehat{f}\left(2n-p\right)T_{\text{s}}+2\widehat{\phi}\right) & =\Re\left(\sum_{n=p+1}^{N}e^{2j\pi\widehat{f}\left(2n-p\right)T_{\text{s}}+2j\widehat{\phi}}\right)\\
 & =\cos\left(2\pi\widehat{f}\left(N+1\right)T_{\text{s}}+2\widehat{\phi}\right)\frac{\sin\left(2\pi\widehat{f}\left(N-p\right)T_{\text{s}}\right)}{\sin\left(2\pi\widehat{f}T_{\text{s}}\right)},
\end{align*}
after some basic calculus, where $\Re\left(z\right)$ is the real
part of $z\in\mathbb{C}$. Finally,
\[
h_{1}\left(p,\widehat{\boldsymbol{\theta}}\right)=\frac{1}{2N}\left(\cos\left(2\pi\widehat{f}\left(N+1\right)T_{\text{s}}+2\widehat{\phi}\right)\frac{\sin\left(2\pi\widehat{f}\left(N-p\right)T_{\text{s}}\right)}{\sin\left(2\pi\widehat{f}T_{\text{s}}\right)}+\left(N-p\right)\cos\left(2\pi\widehat{f}pT_{\text{s}}\right)\right).
\]

One has also to evaluate 
\begin{align*}
\frac{\partial x_{n}^{m}\left(\widehat{\boldsymbol{\theta}}\right)}{\partial\widehat{f}} & =-2\pi nT_{\text{s}}\sin\left(2\pi\widehat{f}nT_{\text{s}}+\widehat{\phi}\right),
\end{align*}
to get
\[
h_{2}\left(p,\widehat{\boldsymbol{\theta}}\right)=\frac{1}{N}\sum_{n=p+1}^{N}4\pi^{2}n\left(n-p\right)T_{\text{s}}^{2}\widehat{a}^{2}\sin\left(2\pi\widehat{f}nT_{\text{s}}+\widehat{\phi}\right)\sin\left(2\pi\widehat{f}\left(n-p\right)T_{\text{s}}+\widehat{\phi}\right)
\]
for which a more compact expression is difficult to get.

Finally, 
\[
\frac{\partial x_{n}^{m}\left(\widehat{\boldsymbol{\theta}}\right)}{\partial\widehat{\phi}}=-\widehat{a}\sin\left(2\pi\widehat{f}nT_{\text{s}}+\widehat{\phi}\right)
\]
leads to
\begin{align*}
h_{3}\left(p,\widehat{\boldsymbol{\theta}}\right) & =\frac{1}{N}\sum_{n=p+1}^{N}\widehat{a}^{2}\sin\left(2\pi\widehat{f}nT_{\text{s}}+\widehat{\phi}\right)\sin\left(2\pi\widehat{f}\left(n-p\right)T_{\text{s}}+\widehat{\phi}\right)\\
 & =\frac{\widehat{a}^{2}}{2N}\left(-\cos\left(2\pi\widehat{f}\left(N+1\right)T_{\text{s}}+2\widehat{\phi}\right)\frac{\sin\left(2\pi\widehat{f}\left(N-p\right)T_{\text{s}}\right)}{\sin\left(2\pi\widehat{f}T_{\text{s}}\right)}+\left(N-p\right)\cos\left(2\pi\widehat{f}pT_{\text{s}}\right)\right),
\end{align*}
obtained via the same derivation as those used to obtain $h_{1}\left(p,\widehat{\boldsymbol{\theta}}\right)$.

\subsection{Auto-correlation function for polynomial model}

\label{subsec:gamma_poly}

Consider the Tchebychev polynomial model of degree $K-1$ with estimated
parameters $\widehat{\theta}_{k}$, $k=1,\dots,K$ given by \eqref{eq:poly}.
Using Proposition~\ref{prop:model_eq2}, one has to evaluate 
\[
\frac{\partial x_{n}^{m}\left(\widehat{\boldsymbol{\theta}}\right)}{\partial\widehat{\theta}_{k}}=\mathcal{T}_{k-1}\left(2\frac{n}{N}-1\right)
\]
to get
\[
h_{k}^{m}\left(p\right)=\frac{1}{N}\sum_{n=p+1}^{N}\mathcal{T}_{k-1}\left(2\frac{n}{N}-1\right)\mathcal{T}_{k-1}\left(2\frac{n-p}{N}-1\right),
\]
independent of $\widehat{\boldsymbol{\theta}}$. 

\subsection{Auto-correlation function for sample predictive model}

\label{subsec:gamma_samples_pred}

Consider the sample predictive coding approach of order $K$, offset
$\eta$, and estimated parameters $\widehat{\theta}_{k}$, $k=1,\dots,K$
given by \eqref{eq:samples_pred}. Using Proposition~\ref{prop:model_eq2},
one has to evaluate
\[
\frac{\partial x_{n}^{m}\left(\widehat{\boldsymbol{\theta}}\right)}{\partial\widehat{\theta}_{k}}=\widehat{x}_{\left(i-1\right)N+n-\eta-k+1}
\]
to get
\[
h_{k}\left(p\right)=\frac{1}{N}\sum_{n=p+1}^{N}\widehat{x}_{\left(i-1\right)N+n-\eta-k+1}\widehat{x}_{\left(i-1\right)N+n-p-\eta-k+1},
\]
independent of $\widehat{\boldsymbol{\theta}}$.
\end{document}